\documentclass[aps,pra,twocolumn,superscriptaddress,longbibliography, 10pt]{revtex4-2}
\usepackage{lmodern}
\usepackage[T1]{fontenc}
\usepackage[utf8]{inputenc}
\usepackage{color}
\usepackage{mathrsfs}
\usepackage{amsmath}
\usepackage{amsthm}
\usepackage{amssymb}
\usepackage{graphicx}
\usepackage{esint}
\PassOptionsToPackage{normalem}{ulem}
\usepackage{ulem}

\usepackage{latexsym}
\usepackage{bm}
\usepackage{amsfonts}
\usepackage{babel}
\usepackage{tikz}
\usepackage{pgfplots}
\usepackage{enumitem}

\usepackage{hyperref}
\hypersetup{
	breaklinks = true,
    colorlinks = true,
    citecolor = {blue},
	urlcolor = {blue},
	linkcolor = {blue}
}

\newcommand{\B}[1]{{\color{blue}#1}}
\newcommand{\R}[1]{{\color{red}#1}}

\makeatother
\allowdisplaybreaks

\usepackage{babel}
\begin{document}
\title{Topological modes and spectral flows in inhomogeneous PT-symmetric continuous media}
\author{Yichen Fu}
\email{fu9@llnl.gov}

\affiliation{Princeton Plasma Physics Laboratory and Department of Astrophysical Sciences, ~\\
 Princeton University, Princeton, NJ 08543, USA}
\affiliation{Lawrence Livermore National Laboratory, Livermore, CA 94550, USA}

\author{Hong Qin}
\email{hongqin@princeton.edu}

\affiliation{Princeton Plasma Physics Laboratory and Department of Astrophysical Sciences, ~\\
 Princeton University, Princeton, NJ 08543, USA}

\begin{abstract}
In classical Hermitian continuous media, the spectral-flow index of topological modes is linked to the bulk topology via index theorem. However, the interface between two bulks is usually non-Hermitian due to the inhomogeneities of system parameters. We show that the connection between topological modes and bulk topology still exists despite the non-Hermiticity at the interface if the system is endowed with PT symmetry. The theoretical framework developed is applied to the Hall magnetohydrodynamic model to identify a topological mode called topological Alfv\'{e}n-sound wave in magnetized plasmas. 
\end{abstract}

\maketitle

\section{Introduction}

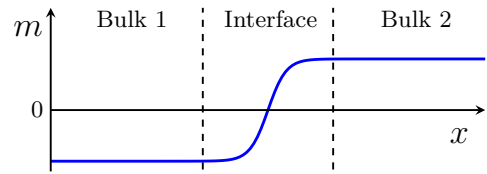
\begin{figure}[ht]
\centering\begin{tikzpicture}[scale=0.9]
            \begin{axis}[
                xmin=0, xmax=4, ymin=-1.2, ymax=2,
                axis lines=center,
                axis on top=true,
                domain=0:4,
                ylabel=$m$, xlabel=$x$,
                xticklabel=\empty, yticklabel=\empty,
                samples=125,
                label style={font=\Large},
                width=8cm, height=4cm,
                xtick style={draw=none},
                ytick style={thick},
                x label style={below right}, y label style={left},
                axis line style = thick,
                ytick={0}
                ]
                \addplot [mark=none, draw=blue, very thick] {tanh(6.0*(\x-2.0))};
                \addplot [mark=none, thick, dashed] coordinates {(2.6, -2) (2.6, 2)};
                \addplot [mark=none, thick, dashed] coordinates {(1.4, -2) (1.4, 2)};
            \end{axis}
    
            \node[align=center, above] at (1.2, 2) {Bulk 1};
            \node[align=center, above] at (5.4, 2) {Bulk 2};
            \node[align=center, above] at (3.25, 2) {Interface};

            \node[align=center, above] at (-0.2, 0.675) {$0$};
    
        \end{tikzpicture} \caption{Sketch of a continuous model. System parameter $m(x)$ is constant in bulk 1 and 2 but changes sign smoothly in the interface.}
\label{fig:configuration}
\end{figure}

Topological modes and classifications of electronic bands in crystal structures \cite{thouless1982quantized,simon1983holonomy,hasan2010colloquium,qi2011topological,armitage2018weyl,kitaev2009periodic,chiu2016classification} have been studied extensively. Similar investigations have been carried out for photonic crystals \cite{haldane2008possible,raghu2008analogs,ozawa2019topological} or phononic crystals \cite{prodan2009topological,zhang2010topological,yang2015topological,wang2015topological}. Topological analysis has also found applications in continuous models, which can describe either the linear waves in classical continuous media, such as neutral fluids \cite{souslov2019topological,delplace2017topological,perrot2019topological,perez2022unidirectional,Zhu2023,tong2023gauge} and plasmas \cite{gao2016photonic,parker2020topological,fu2021topological,fu2022dispersion,palmerduca2023photon}, or low-energy approximation of some electronic bands \cite{bernevig2006quantum,zyuzin2012topological,leykam2017edge,shen2018topological}. In 2D Hermitian cases, a topological mode in continuous models can be understood as spectral flow induced by the nontrivial topology of the eigenmode bundle over a sphere surrounding the phase-space Weyl point \cite{delplace2022berry,qin2022topological,faure2023manifestation,marciani2020chiral}. A typical continuous model, which admits topological modes, is sketched in Fig.~\ref{fig:configuration}, where a system parameter $m$ is constant in two bulk regions but changes sign in the interface region. If the system is Hermitian and supports a Weyl point, then according to the index theorem \cite{faure2023manifestation}, the spectral-flow index in the band gap equals the topological charge of the Weyl point.

Recently, non-Hermitian extensions \cite{bergholtz2021exceptional,wang2021topological,okuma2023non} of those topological properties have attracted considerable attention. The bulk-boundary correspondence was found to be violated in non-Hermitian systems \cite{lee2016anomalous,xiong2018does} due to the non-Hermitian skin effect \cite{yao2018edge,yao2018non,song2019non,okuma2020topological}. Various strategies have been developed to reestablish the non-Hermitian bulk-boundary correspondence, including defining the generalized Brillouin zone \cite{yao2018edge} and applying biorthogonal formalism \cite{kunst2018biorthogonal}. Non-Hermitian phenomena were discovered without Hermitian counterparts, such as the exceptional degeneracies \cite{berry2004physics,heiss2012physics} and the distinction between point and line gaps \cite{kawabata2019classification}. The classification of gapless and gapped non-Hermitian systems was also proposed \cite{gong2018topological,kawabata2019classification,kawabata2019symmetry}. In continuous media, although such bulk-edge correspondence was recently demonstrated for a class of designed non-Hermitian systems that is deformed from Hermitian systems and preserve the point degeneracy and a line gap \cite{jezequel2023non}, the general properties of non-Hermitian spectral flows in continuous models still need to be further investigated.

Instead of attempting a general description of non-Hermitian spectral flows, we focus on a special class of problems that naturally arise and are typical in classical continuous media. Classical homogeneous media are Hermitian in general, but classical inhomogeneous media are mostly non-Hermitian. For the setup shown in Fig.~\ref{fig:configuration}, when $m(x)$ represents the flow speed in fluids and plasmas, the two homogeneous bulk regions are Hermitian, but the interface region is non-Hermitian due to the flow shear, which supports the Kelvin-Helmholtz instability under appropriate conditions. For this system, we ask the following question: Can the topology from the Hermitian bulk predict the spectral flows in the non-Hermitian interface? Although it may not be possible in general without additional structures, we show that the answer to this question is affirmative when the system is endowed with Parity-Time (PT) symmetry \cite{bender1998real,bender2007making,el2018non,Zhang2020} and the non-Hermiticity is relatively weak, and in this case the corresponding spectral flows are stable surface waves with real eigenfrequencies. It turns out that PT symmetry is a general property that appears in many classical non-dissipative systems \cite{Qin2021}, including the shear flow system above \cite{qin2019kelvin,fu2020physics}, or system with a balance of loss and gain \cite{ozdemir2019parity,miri2019exceptional}.

We start from a family of simple two-band Hamiltonians to demonstrate the key physics of topological edge modes, spectrum flows, and topological charges in inhomogeneous continuous systems. The general formalism for non-Hermitian inhomogeneous continuous systems beyond the two-band model is then established, which is applied to identify a topological edge mode called topological Alfv\'{e}n-sound wave (TASW) in magnetized plasmas.

\section{A two-band model} 

Before presenting the result for a general inhomogeneous PT-symmetric continuous media, we consider the following two-band non-Hermitian Hamiltonians as simple examples of asymptotically Hermitian systems, 
\begin{align}
\mathcal{H}_{\epsilon}^{\alpha}[\lambda;x,\partial_{x}]=\begin{pmatrix}m(x) & \lambda-\partial_{x}\\
\lambda+\partial_{x} & -m(x)
\end{pmatrix}+\mathrm{i}\epsilon m'(x)\sigma^{\alpha}.\label{eq:Hoperator2x2}
\end{align}
Here, $m(x)$ is a real function of $x$, representing a system parameter, such as the mass term, and $m'\doteq\mathrm{d}m/\mathrm{d}x$. $\lambda$ is a control parameter, $\sigma^{\alpha}(\alpha=0,x,y,z)$ is an identity matrix or one of the Pauli matrices, and $\epsilon$ is a label indicating the order of the anti-Hermitian term. To simplify the presentation, we define the notion of asymptotic Hamiltonian. For a given Hamiltonian, when all derivatives of the system parameters are set to zero, it will be called the asymptotic Hamiltonian of the original Hamiltonian. An inhomogeneous non-Hermitian continuous system is called asymptotically Hermitian if its asymptotic Hamiltonian is Hermitian.

To study the possible topological modes of $\mathcal{H}_{\epsilon}^{\alpha}[\lambda;x,\partial_{x}]$, we need to investigate the corresponding symbol defined by the Wigner-Weyl transform $H_{\epsilon}^{\alpha}=\mathscr{W}(\mathcal{H}_{\epsilon}^{\alpha})$ \cite{moyal1949quantum,hillery1984distribution}. For the simple Hamiltonian in the form of Eq.\,(\ref{eq:Hoperator2x2}), its symbol can be obtained by replacing $\hat{k}\doteq-\mathrm{i}\partial_{x}$ with $k$, 
\begin{align}
H_{\epsilon}^{\alpha}[\lambda;x,k]=\begin{pmatrix}m(x) & \lambda-\mathrm{i}k\\
\lambda+\mathrm{i}k & -m(x)
\end{pmatrix}+\mathrm{i}\epsilon m'\sigma^{\alpha}.\label{eq:2_by_2_demo}
\end{align}
The asymptotic Hamiltonian $H_{\epsilon=0}^{\alpha}$ is Hermitian with eigenvalues $\omega^{\pm}=\pm\sqrt{m^{2}+k^{2}+\lambda^{2}}$. Thus, there is a Weyl point (band crossing point) at $(m,k,\lambda)=(0,0,0)$, and its topological index can be easily calculated \cite{delplace2022berry,qin2022topological,faure2023manifestation}. $\mathcal{H}_{\epsilon=0}^{\alpha}$ is known as a Dirac operator that supports a spectral flow with respect to the control parameter $\lambda$.

When $\epsilon>0$, in general, $H_{\epsilon}^{\alpha}$ is non-Hermitian with complex eigenvalues, supports exceptional degeneracies \cite{xu2017weyl,cerjan2018effects,cerjan2019experimental}, and does not have a line gap. When $\alpha\in\{0,x,z\}$ the spectrum of $\mathcal{H}_{\epsilon}^{\alpha}[\lambda;x,\partial_{x}]$ is also complex. Therefore, we need to explore the non-Hermitian topology emergent from the exceptional degeneracies in the interface region to fully understand their spectral flow. However, we observe that $\mathcal{H}_{\epsilon}^{y}$ is an exception, which has a real spectrum for all $\epsilon\in\mathbb{R}$. Its topology can be captured from the Weyl point determined by its asymptotic Hamiltonian.

Here, we show why $\mathcal{H}_{\epsilon}^{y}$ is an exception from two perspectives. Firstly, $(\mathcal{H}_{\epsilon}^{y})^{*}=\mathcal{H}_{\epsilon}^{y}$, i.e., it is a real operator. Its anti-Hermitian part $\mathrm{i}\epsilon m'(x)\sigma^{y}$ can be combined with the $\partial_{x}$ operator as a ``covariant derivative'' $(\partial_{x}-\epsilon m')$. To calculate its spectrum, we perform a similarity transformation, 
\begin{align}
\mathcal{H}_{\epsilon}^{y}\to\mathcal{\widetilde{H}}_{\epsilon}^{y}=e^{-\epsilon m}\mathcal{H}_{\epsilon}^{y}e^{\epsilon m}=\mathcal{H}_{0}^{y}.\label{eq:transform_anti_hermitian}
\end{align}
The non-Hermitian operator $\mathcal{H}_{\epsilon}^{y}$ has the same spectrum as the Hermitian operator $\mathcal{H}_{0}^{y}$. In particular, the $\mathcal{H}_{\epsilon}^{y}$ will have the same topological modes and spectral flows as $\mathcal{H}_{0}^{y},$ which is fully determined by its symbol $H_{0}^{y}=\mathscr{W}(\mathcal{H}_{0}^{y})$. In addition, eigenvectors are transformed as $\tilde{\Psi}\to\mathrm{e}^{-\epsilon m}\Psi$, which resembles the non-Hermitian skin effect in lattices with open boundaries \cite{yao2018edge}. 

Next, we interpret this result as a consequence of the non-Hermitian Hamiltonian being PT-symmetric. In the current context, time-reversal $\mathcal{T}$ is complex conjugation, and parity $\mathcal{P}$ is a $2\times2$ constant unitary matrix, which satisfies $\mathcal{P}^{2}=1$ and $[\mathcal{P},\mathcal{T}]=0$. It is straightforward to verify that $\mathcal{H}_{\epsilon}^{y}$ is PT-symmetric with $\mathcal{P}=\sigma^{0}$, which implies that $\mathcal{H}_{\epsilon}^{y}$ can have real spectrum despite being non-Hermitian, when PT symmetry is not spontaneously broken \cite{bender2007making,beekman2019introduction}. Since $\mathcal{H}_{\epsilon}^{y}$ is also asymptotically Hermitian, the domain where PT symmetry is not spontaneously broken must contain a neighborhood of $\epsilon=0$. It turns out that this domain of unbroken PT symmetry includes all $\epsilon\in\mathbb{R}$, as implied by Eq.\,(\ref{eq:transform_anti_hermitian}).

The important fact that $\mathcal{H}_{\epsilon}^{y}$ is similar to $\mathcal{H}_{0}^{y}$ for all $\epsilon\in\mathbb{R}$ can be generalized to the following class of generic two-band operators, 
\begin{align}
\mathcal{H}[\lambda;x,\partial_{x}]=\Big[\lambda{d}_{1\mu}+m{d}_{2\mu}-\mathrm{i}\partial_{x}{d}_{3\mu}+\mathrm{i}nd_{4\mu}\Big]{\sigma}^{\mu},\label{eq:generic_two_band}
\end{align}
where $m(x)\,\text{and }n(x)$ are two real functions of $x$, $\sigma^{\mu}=(\sigma^{0},\boldsymbol{\sigma})$ is Pauli 4-vector, $d_{i\mu}$ are constant real 4-vectors. The Hamiltonian in Eq.\,(\ref{eq:generic_two_band}) is assumed to be PT-symmetric for a properly chosen $\mathcal{P}.$ We prove that $\mathcal{H}[\lambda;x,\partial_{x}]$ is similar to a Hermitian operator.

The Hermitian part of its symbol $H_{\mathrm{H}}[\lambda;x,k]=[\lambda{d}_{1\mu}+m(x){d}_{2\mu}+k{d}_{3\mu}]{\sigma}^{\mu}$ supports a degeneracy point at $(\lambda,m,k)=(0,0,0)$. $\mathcal{H}$ being PT-symmetric means 
\begin{align}
(\mathcal{PT})\mathcal{H}(\mathcal{TP})=\mathcal{P}\mathcal{H}^{*}\mathcal{P}=\mathcal{H.}\label{eq:PT_property1}
\end{align}
Pick two orthonormal eigenvectors $\Psi_{i=1,2}$ of $\mathcal{PT}$ with eigenvalues $1$\footnote{Some antilinear operators do not have eigenvectors. For instance, with $\mathcal{P}=\mathrm{i}\sigma^{y}$, $\mathcal{PT}$ has no eigenvector because $(\mathcal{PT})^{2}=-1$. However, since classical systems have $(\mathcal{PT})^{2}=1$, this issue is avoided. See Ref.~\cite{uhlmann2016anti} for example.}, $\mathcal{PT}\Psi_{i}=\mathcal{P}\Psi_{i}^{*}=\Psi_{i}$. Here, $\Psi_{i}$ are constant vectors. Let $\boldsymbol{\Psi}\doteq(\Psi_{1},\Psi_{2})^{\intercal}$, and it induces a similarity transformation for $\mathcal{H}$, $\mathcal{H}'=\boldsymbol{\Psi}^{\dagger}\mathcal{H}\boldsymbol{\Psi}$. Let $\boldsymbol{\Psi}^{\dagger}\sigma^{\mu}\boldsymbol{\Psi}=c_{\nu}^{\mu}\sigma^{\nu}$, we find $\mathcal{H}'$ has the same form as Eq.~(\ref{eq:generic_two_band}) with $d_{i\mu}$ replaced by $\tilde{d}_{i\mu}\doteq c_{\mu}^{\nu}d_{i\nu}$. Since $\boldsymbol{\Psi}^{\dagger}\sigma^{\mu}\boldsymbol{\Psi}$ is Hermitian, $c_{\nu}^{\mu}$ must be real, so are $\tilde{d}_{i\mu}$.

From Eq.~(\ref{eq:PT_property1}), we have 
\begin{equation}
\begin{aligned}(\mathcal{H}_{ij}')^{*} & =(\Psi_{i}^{*})^{\dagger}\mathcal{H}^{*}\Psi_{j}^{*}=(\mathcal{P}\Psi_{i})^{\dagger}(\mathcal{P}\mathcal{H}\mathcal{P})(\mathcal{P}\Psi_{j})\\
 & =\Psi_{i}^{\dagger}\mathcal{P}^{\dagger}\mathcal{P}\mathcal{H}\mathcal{P}^{\dagger}\mathcal{P}\Psi_{j}=\mathcal{H}_{ij}',
\end{aligned}
\label{eq:prove_of_real}
\end{equation}
which means that $\mathcal{H}'$ is a real operator. Thus, the coefficients in front of $\sigma_{y}$ must be imaginary, and the coefficients of $\sigma_{0},\sigma_{x},\sigma_{z}$ must be real. So $\mathcal{H}'$ has the following form, 
\begin{align}
\mathcal{H}'=\sum_{\mu}^{0,x,z}\Big[\lambda\tilde{d}_{1\mu}+m\tilde{d}_{2\mu}\Big]{\sigma}^{\mu}+\mathrm{i}\Big[n\tilde{d}_{4y}-\partial_{x}\tilde{d}_{3y}\Big]{\sigma}^{y}.
\end{align}
Let $\gamma\doteq\int n(x)\tilde{d}_{4y}/\tilde{d}_{3y}\,\mathrm{d}x$. The anti-Hermitian part of $\mathcal{H}'$ can be transformed away by a similarity transform, 
\begin{equation}
\mathcal{H}''=\mathrm{e}^{-\gamma}\mathcal{H}'\mathrm{e}^{\gamma}=\sum_{\mu}^{0,x,z}\Big[\lambda\tilde{d}_{1\mu}+m\tilde{d}_{2\mu}\Big]{\sigma}^{\mu}-\mathrm{i}\partial_{x}\tilde{d}_{3y}{\sigma}^{y}.\label{eq:H''}
\end{equation}
The spectrum of the PT-symmetric $\mathcal{H}$ in Eq.~(\ref{eq:generic_two_band}) is identical to that of the Hermitian operator $\mathcal{H}''$. In particular, $\mathcal{H}$ admits topological edge modes whose spectral-flow index is determined by the symbol $H''=\mathscr{W}(\mathcal{H}'')$ according to Faure's index theorem \cite{faure2023manifestation}.

\section{Inhomogeneous PT-symmetric continuous media} 

We now further generalize the result by showing that when an inhomogeneous medium is asymptotically Hermitian and PT-symmetric, and if the asymptotic Hamiltonian has a band gap near a Weyl point of two-fold degeneracy\footnote{It is equivalent to restricting our discussion on type-I Weyl points defined in Ref.\cite{soluyanov2015type}.}, then the non-Hermitian inhomogeneous system admits topological edge modes with real frequency in the neighborhood of the Weyl point. The spectral-flow index of the mode is identical to the topological charge of the Weyl point. This result is established by proving that the system Hamiltonian can be approximated by a two-band Hamiltonian of the form of Eq.\,(\ref{eq:generic_two_band}) near the Weyl point of two-fold degeneracy of the asymptotic Hamiltonian symbol. 

Consider a 1D continuous media with $N$ field variables. The dynamics of the system is governed by an asymptotically Hermitian, PT-symmetric Hamiltonian operator $\mathcal{H}[\lambda;x,\partial_{x}]$, which depends on coordinate $x$ through a system parameter $m(x)$ and its derivatives. Here, we make a technical assumption that the inhomogeneity experienced by the edge mode is weak, i.e., $\epsilon\doteq\delta/l\ll1$, where $\delta$ is the scale length of the edge mode and $l$ that of $m(x)$. In dimensionless variable, it implies that the $i$-th order derivatives $m^{(i)}$ is of order $O(\epsilon^{i})$. This technical assumption of weak inhomogeneity is consistent with asymptotic Hermiticity. We write $\mathcal{H}$ as 
\begin{equation}
\begin{aligned}\mathcal{H}= & \mathcal{H}_{0}[\lambda;m,\partial_{x}]+\epsilon\mathcal{H}_{1}[\lambda;m,m',\partial_{x}]\\
 & +\epsilon^{2}\mathcal{H}_{2}[\lambda;m,m',m'',\partial_{x}]+\cdots,
\end{aligned}
\label{eq:operator_expression}
\end{equation}
where $\mathcal{H}_{i}$ depends on up to the $i$-th order derivatives $m^{(i)}$, and $\epsilon^{i}$ is a label indicating that $\mathcal{H}_{i}$ is $O(\epsilon^{i})$. Similarly, the symbol $H[\lambda;x,k]$ can be written as 
\begin{equation}
\begin{aligned}H\doteq\mathscr{W}[\mathcal{H}]= & H_{0}[\lambda;m,k]+\epsilon H_{1}[\lambda;m,m',k]\\
 & +\epsilon^{2}H_{2}[\lambda;m,m',m'',k]+\cdots,
\end{aligned}
\label{eq:symbol_expression}
\end{equation}
where $\epsilon^{i}$ is a label indicating that $H_{i}$ is $O(\epsilon^{i})$ when $k$ is taken to be $O(\epsilon).$ Note that due to the inhomogeneity, $H_{i}\neq\mathscr{W}[\mathcal{H}_{i}]$ in general. By the assumption of asymptotic Hermiticity, $\mathcal{H}_{0}$ and $H_{0}$ are Hermitian at each value of $x$, and $\mathcal{H}_{i\geq1}$ and $H_{i\geq1}$ are allowed to be non-Hermitian due to the inhomogeneity. The expansion of the Hamiltonian according to the weak inhomogeneity and the assumption of asymptotic Hermiticity requires that non-Hermiticity and the free energy responsible for possible instabilities can only be placed at $H_{i\geq1}$ and $\mathcal{H}_{i\geq1}$.

We will show that under the assumptions of asymptotic Hermiticity and PT symmetry, the spectral flow of the non-Hermitian $\mathcal{H}[\lambda;x,\partial_{x}]$ will be determined by the asymptotic symbol $H_{0}[\lambda;m,k]$. For convenience, we call $H_{0}[\lambda;m,k]$ the bulk symbol and its eigenvalues $\omega_{n}(\lambda;m,k)$ the bulk bands at a given value of $m(x)$ determined by a given $x$. Without losing generality, let $(\lambda,x,k)=(0,0,0)$ be an isolated Weyl point of two-fold degeneracy of the bulk bands. If $\mathcal{H}$ is Hermitian and 
\begin{align}
H\doteq\mathscr{W}[\mathcal{H}]=H_{0}[\lambda;m,k],\label{eq:H=00003D00003DDH0}
\end{align}
then spectral-flow index of the edge modes of $\mathcal{H}$ near the Weyl point is determined by $H_{0}$ according to Faure's index theorem \cite{faure2023manifestation}. In certain simple inhomogeneous systems, such as those studied in Ref.~\cite{souslov2019topological,delplace2017topological,delplace2022berry,parker2020topological,fu2021topological}, the conditions of $\mathcal{H}$ being Hermitian and Eq. (\ref{eq:H=00003D00003DDH0}) are satisfied, but in general they are not. Nevertheless, we show that when an inhomogeneous $\mathcal{H}$ is asymptotically Hermitian and PT-symmetric, $\mathcal{H}$ still admits topological edge modes in the neighborhood of the Weyl point of $H_{0}$, and the spectral-flow index of $\mathcal{H}$ is also determined by the topology of $H_{0}$.

The definitions of $\mathcal{P}$ and $\mathcal{T}$ operators are similar to those for the two-band systems discussed above, except that $\mathcal{P}$ is now an $N\times N$ constant matrix. From Eqs.~(\ref{eq:operator_expression}) and (\ref{eq:symbol_expression}), $\mathcal{H}_{i}$ and $H_{i}$ are PT-symmetric for all $i\geq0$.

Assume the asymptotic symbol $H_{0}[\lambda;m,k]$ has a Weyl point of two-fold degeneracy at $(\lambda,x,k)=(0,0,0)$, its eigenvalues are $\omega_{1}=\omega_{2}=0$ with eigenvectors $\Psi_{1}$ and $\Psi_{2}$. To study the behavior of operator $\mathcal{H}$ near the Weyl point, we expand the symbol $H[\lambda;x,k]$ up to $O(\epsilon)$, 
\begin{equation}
\begin{aligned}H\approx & \tilde{H}[\lambda;x,k]\\
\doteq & H_{0}[0;0;0]+\lambda\dfrac{\partial H_{0}}{\partial\lambda}+m'(0)x\dfrac{\partial H_{0}}{\partial m}+k\dfrac{\partial H_{0}}{\partial k}\\
 & +H_{1}[0;0,m'(0),0].
\end{aligned}
\end{equation}
Because $H_{0}$, $\partial H_{0}$ and $H_{1}$ are all evaluated at the degeneracy point, they do not depend on $\lambda,x$ or $k$. Therefore, the corresponding approximated operator is 
\begin{align}
\tilde{\mathcal{H}}\doteq & \mathscr{W}^{-1}[\tilde{H}]\nonumber \\
= & H_{0}+\lambda\dfrac{\partial H_{0}}{\partial\lambda}+m'(0)x\dfrac{\partial H_{0}}{\partial m}-\mathrm{i}\partial_{x}\dfrac{\partial H_{0}}{\partial k}+H_{1}.
\end{align}
Near the Weyl point, we can focus on the dynamics in the subspace of degeneracy and approximate $\tilde{\mathcal{H}}$ by a $2\times2$ operator, 
\begin{align}
 & \mathcal{M}[\lambda;m(x),m'(x=0),\partial_{x}]\doteq\boldsymbol{\Psi}^{\dagger}\tilde{\mathcal{H}}\boldsymbol{\Psi}\nonumber \\
= & \boldsymbol{\Psi}^{\dagger}\left(\lambda\dfrac{\partial H_{0}}{\partial\lambda}{+}m'x\dfrac{\partial H_{0}}{\partial m}{-}\mathrm{i}\partial_{x}\dfrac{\partial H_{0}}{\partial k}\right)\boldsymbol{\Psi}+\boldsymbol{\Psi}^{\dagger}H_{1}\boldsymbol{\Psi}.\label{eq:reduced_2_by_2}
\end{align}
Here, $\boldsymbol{\Psi}\doteq(\Psi_{1},\Psi_{2})^{\intercal}$. Since $H_{0}$ is Hermitian, so are $\partial H_{0}$ and $\boldsymbol{\Psi}^{\dagger}\partial H_{0}\boldsymbol{\Psi}$. In contrast, $\boldsymbol{\Psi}^{\dagger}H_{1}\boldsymbol{\Psi}$ could have both Hermitian and anti-Hermitian parts. At this point, we recognize that operator $\mathcal{M}$ assumes the form of the two-band Hamiltonian in Eq.~(\ref{eq:generic_two_band}).

Now we invoke the PT-symmetry condition. Since $H_{1}$ and $H_{2}$ are PT-symmetric, so are $\tilde{H}$ and $\tilde{\mathcal{H}}$. Because $\Psi_{1}$ and $\Psi_{2}$ are eigenvectors of $H_{0}$ with real eigenvalues, they are also eigenvectors of the $\mathcal{PT}$ operator. In a manner similar to Eq.~(\ref{eq:prove_of_real}), it's easy to show that $\mathcal{M}$ is a real operator. The coefficient in front of $\sigma_{y}$ must be imaginary, and the anti-Hermitian part $\mathcal{M_{\textrm{A}}}$ can be absorbed by a similarity transformation, and do not affect the operator's spectrum, as in Eq.\,(\ref{eq:H''}). The topological edge modes and spectral flows of $\mathcal{M}$ are the same as those of its Hermitian part $\mathcal{M}_{\textrm{H}},$ whose spectral-flow index is determined by $M_{\textrm{H}}=\mathscr{W}[\mathcal{M}_{\textrm{H}}].$ As the small contribution of $\boldsymbol{\Psi}^{\dagger}H_{1}\boldsymbol{\Psi}$ in $M_{\textrm{H}}$ is a constant Hermitian matrix, $M_{\textrm{H}}$ only perturbs the location of the Weyl point and band gap of $H_{0}$ without changing the topology. This concludes our proof that when an inhomogeneous $\mathcal{H}$ is asymptotically Hermitian and PT-symmetric, $\mathcal{H}$ admits topological edge modes in the neighborhood of the Weyl point of $H_{0}$, and the spectral-flow index of $\mathcal{H}$ is also determined by $H_{0}$. Since the eigenvalues are real, such topological edge modes are stable.

\section{Topological Alfv\'{e}n-sound wave} 

We now apply the general theoretical framework developed to identify a topological edge mode called topological Alfv\'{e}n-sound wave in the Hall magnetohydrodynamics (MHD) model, which describes low-frequency, long-wavelength dynamics of magnetized plasmas \cite{goedbloed_poedts_2004,freidberg_2014,hameiri2004linear,hameiri2005waves}.

We study the linearized waves in an inhomogeneous 1D Hall MHD equilibrium. The plasma is assumed to be adiabatic, namely, $p/\rho^{\gamma}$ is constant, where $p$ is plasma pressure, $\rho$ is plasma density, and $\gamma$ is the ratio of specific heats. Consider an equilibrium with constant density and no mass flow. The equilibrium magnetic field $\mathbf{B}$ and pressure $p$ are balanced according to 
\begin{align}
\dfrac{1}{\mu_{0}}(\nabla\times\mathbf{B})\times\mathbf{B}=\nabla p,\label{eq:equilibrium}
\end{align}
where $\mu_{0}$ is the vacuum permeability. The perturbed field $\Psi=(\tilde{\mathbf{v}},\tilde{\mathbf{B}},\tilde{p})^{\intercal}$ relative to the equilibrium evolve according to the linearized system (See Appendix~\ref{sec:hmhd_equations}), 
\begin{align}
\partial_{t}\tilde{\mathbf{v}}= & (\nabla{\times}\tilde{\mathbf{B}})\times\mathbf{v}_{\mathrm{A}}+(\nabla{\times}\mathbf{v}_{\mathrm{A}})\times\tilde{\mathbf{B}}-\nabla(v_{\mathrm{s}}\tilde{p}),\label{eq:linearized_momentum}\\[3pt]
\partial_{t}\tilde{\mathbf{B}}= & \nabla\times(\tilde{\mathbf{v}}\times\mathbf{v}_{\mathrm{A}})\nonumber \\
 & -d_{\mathrm{i}}\nabla{\times}\left[(\nabla{\times}\tilde{\mathbf{B}})\times\mathbf{v}_{\mathrm{A}}+(\nabla{\times}\mathbf{v}_{\mathrm{A}})\times\tilde{\mathbf{B}}\right],\label{eq:linearized_induction}\\[3pt]
\partial_{t}\tilde{p}= & -v_{\mathrm{s}}\nabla{\cdot}\tilde{\mathbf{v}}-(2/\gamma)\tilde{\mathbf{v}}{\cdot}\nabla v_{\mathrm{s}},\label{eq:linearized_state}
\end{align}
where $\mathbf{v}_{\mathrm{A}}\doteq\mathbf{B}/\sqrt{\mu_{0}\rho}$ is the Alfv\'{e}n velocity, $v_{\mathrm{s}}\doteq\sqrt{\gamma p/\rho}$ is the sound velocity, and $d_{\mathrm{i}}$ is the ion skin depth, which is a constant controlling the magnitude of the Hall effect. Equations (\ref{eq:linearized_momentum})-(\ref{eq:linearized_state}) can be cast into a Schrödinger equation 
\begin{align}
\mathrm{i}\partial_{t}\Psi & =\mathcal{H}\Psi=(\mathcal{H}_{0}+\mathcal{H}_{1}+\mathcal{H}_{2})\Psi,\label{eq:operator_decomposition}\\
\mathcal{H}_{0} & =\begin{pmatrix}0 & (\mathbf{v}_{\mathrm{A}}\hat{\mathbf{k}})^{\intercal}-\mathbf{v}_{\mathrm{A}}{\cdot}\hat{\mathbf{k}}\,\, & v_{\mathrm{s}}\hat{\mathbf{k}}\\[3pt]
(\mathbf{v}_{\mathrm{A}}\hat{\mathbf{k}})-\mathbf{v}_{\mathrm{A}}{\cdot}\hat{\mathbf{k}}\,\, & \mathrm{i}d_{\mathrm{i}}(\mathbf{v}_{\mathrm{A}}{\cdot}\hat{\mathbf{k}})\hat{\mathbf{k}}\times & 0\\[3pt]
v_{\mathrm{s}}\hat{\mathbf{k}}^{\intercal} & 0 & 0
\end{pmatrix}.\label{eq:operator_H0}
\end{align}
Here, $\mathcal{H}_{0}$ is the asymptotic Hamiltonian that does not depend on the spatial derivatives of the equilibrium field, and $\mathcal{H}_{1}$ and $\mathcal{H}_{2}$ depends on the first and second order derivatives, respectively. The expressions of $\mathcal{H}_{1}$ and $\mathcal{H}_{2}$ are given in Appendix~\ref{sec:TAS}. The corresponding symbol $H=H_{0}+H_{1}+H_{2}$ also has three parts, where $H_{0}$ can be obtained from $\mathcal{H}_{0}$ by simply replacing $\hat{\mathbf{k}}\to\mathbf{k}$. $\mathcal{H}_{0}$ and $H_{0}$ are Hermitian, $\mathcal{H}$ and $H$ are not in a general equilibrium.

We assume the equilibrium is inhomogeneous only in the $x$-direction, and the background magnetic field is in the $z$-direction, so $\mathbf{v}_{\mathrm{A}}=v_{\mathrm{A}}\mathbf{e}_{z}$. The equilibrium condition given by Eq.~(\ref{eq:equilibrium}) reduces to 
\begin{align}
\dfrac{v_{\mathrm{A}}^{2}(x)}{2}+\dfrac{v_{\mathrm{s}}^{2}(x)}{\gamma}=\mathrm{Const.}\label{eq:equilibrium_constrain}
\end{align}

Let $k_{z}$ be a constant and consider $k_{y}$ as a control parameter. The eigenvalues $\omega$ of $H_{0}$ satisfy the dispersion relation 
\begin{align}
(\omega^{2}{-}k_{z}^{2}v_{\mathrm{A}}^{2})^{2}-(d_{\mathrm{i}}v_{\mathrm{A}}k_{z}k\omega)^{2}=\dfrac{\omega^{2}{-}k_{z}^{2}v_{\mathrm{A}}^{2}}{\omega^{2}{-}k^{2}v_{\mathrm{s}}^{2}}\,k_{\perp}^{2}v_{\mathrm{A}}^{2}\omega^{2},\label{eq:bulk_dispersion}
\end{align}
where $k_{\perp}^{2}=k_{x}^{2}+k_{y}^{2}$ and $k^{2}=k_{\perp}^{2}+k_{z}^{2}$. Since it is derived from $H_{0}$, the dispersion relation does not depend on any derivatives of the equilibrium fields. For $v_{\mathrm{A}}>v_{\mathrm{s}}>0$, there is a Weyl point of two-fold degeneracy between the Alfv\'{e}n wave and sound wave in the positive-frequency branches at $(k_{y},k_{x})=(0,0)$ and 
\begin{align}
\zeta\doteq\dfrac{v_{\mathrm{s}}}{v_{\mathrm{A}}}=\sqrt{1+\dfrac{1}{4}k_{z}^{2}d_{\mathrm{i}}^{2}}-\dfrac{1}{2}k_{z}d_{\mathrm{i}}.\label{eq:speed_ratio}
\end{align}
See Appendix~\ref{sec:hmhd_equations} for details. The question to answer is whether there is a spectral flow of $\mathcal{H}$ in the band gap. Recall that $\mathcal{H}$ is not Hermitian and $\mathscr{W}[\mathcal{H}]=H\neq H_{0}$ for the inhomogeneous equilibrium, and Faure's index theorem \cite{faure2023manifestation} is not applicable. But, with a 1D equilibrium profile satisfying Eq.~(\ref{eq:equilibrium_constrain}), we found that the operator $\mathcal{H}[k_{y};x,\partial_{x}]$ in Eq.~(\ref{eq:operator_decomposition}) is PT-symmetric for $\mathcal{P}=\mathrm{diag}(-1,1,1,-1,1,1,1)$, in addition to be asymptotically Hermitian. (See Appendix~\ref{sec:TAS}). Therefore, based on the analysis established above, we shall expect a spectral flow in $\mathcal{H}$ around the Weyl point. Because the Weyl point and band gap are created by the resonance (band-crossing) between the Alfv\'{e}n wave and sound wave, we will call the spectral flow topological Alfv\'{e}n-sound wave (TASW).

\begin{figure}[ht]
\centering 
\includegraphics[width=8cm]{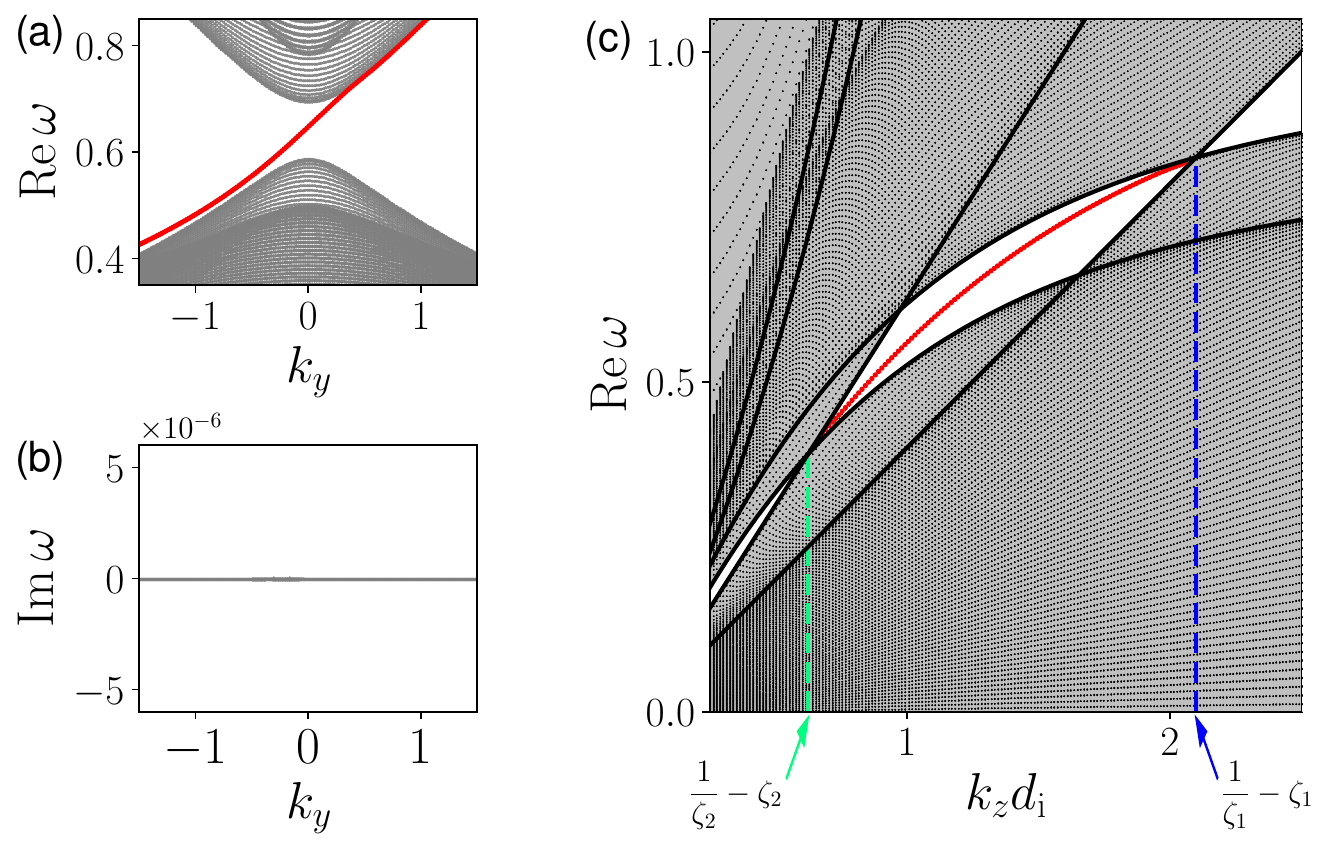} 
\caption{The numerically calculated spectrum of operator $\mathcal{H}$. When $k_{z}=1.25$, the real and imaginary parts of the spectrum as functions of $k_{y}$ are shown in (a) and (b), respectively. The spectral flow, i.e., the TASW, is depicted in red, and the bulk modes are gray. When $k_{y}=0$, the real part of the spectrum as a function of $k_{z}$ is shown in (c). For clarity, the bulk region is filled in gray, while the frequency gap of bulk waves is filled in white. The black lines represent the dispersion relations in two bulk regions calculated from Eq.~(\ref{eq:bulk_dispersion}) at $k_{\perp}=0$. The locations of two degeneracy points are labeled in green and blue dashed lines. }
\label{fig:hmhd_no_shear}
\end{figure}

To verify the existence of TASW, we study the full spectrum of $\mathcal{H}$ numerically. In a 1D domain $[0,L]$, let $v_{\mathrm{A}}(x)=v_{\mathrm{A2}}+0.5(v_{\mathrm{A1}}-v_{\mathrm{A2}})[1+\tanh(\frac{L/2-x}{l})]$, where $v_{\mathrm{A1}},v_{\mathrm{A2}}$ are the Alfv\'{e}n velocities in the two bulk regions, $L$ is the system size, $l$ controls the width of the interface, and $v_{\mathrm{s}}$ is determined from Eq.~(\ref{eq:equilibrium_constrain}). As an example, we choose $v_{\mathrm{A1}}=1,v_{\mathrm{A2}}=0.85,v_{\mathrm{s1}}=0.4,v_{\mathrm{s2}}=0.625,L=100,l=2,d_{\mathrm{i}}=1,$ and $k_{z}=1.25$. The Weyl point is located at $\zeta=0.55$. We expect a degeneracy point somewhere in the interface region because in the two bulk regions $\zeta_{1}=0.4$, $\zeta_{2}=0.73$. The numerical scheme to calculate the operator's spectrum $\mathcal{H}$ is similar to that in Ref.~\cite{fu2022dispersion}. The real part of the spectrum displayed in Fig.~\ref{fig:hmhd_no_shear}(a) clearly shows a spectral flow, i.e., the TASW, and Fig.~\ref{fig:hmhd_no_shear}(b) confirms that the spectrum is real ($|\mathrm{Im}\,\omega|<10^{-8}$). These numerical results agree with our theoretical prediction developed above. The TASW can also be faithfully modeled by a titled Dirac cone \cite{qin2022topological}(see Appendix~\ref{sec:TAS}).

For a given equilibrium profile with fixed $\zeta_{1}=v_{\mathrm{s1}}/v_{\mathrm{A1}}$ and $\zeta_{2}=v_{\mathrm{s2}}/v_{\mathrm{A2}}$, Eq.~(\ref{eq:speed_ratio}) can be solved to find the following range of $k_{z}$ where the TASW exists, 
\begin{align}
\dfrac{1}{\zeta_{2}}-\zeta_{2}<k_{z}d_{\mathrm{i}}<\dfrac{1}{\zeta_{1}}-\zeta_{1}.\label{eq:kz_condition}
\end{align}
In Fig.~\ref{fig:hmhd_no_shear}(c), this condition is numerically verified by the numerically calculated spectrum of $\mathcal{H}$ at different $k_{z}$ with $k_{y}=0$. A frequency gap exists at all $k_{z}$ except for the Weyl points. However, the spectral flow, which is depicted in red, only shows up in the region where condition~(\ref{eq:kz_condition}) is satisfied. It resembles the Fermi arc in topological electronic crystals. These numerical results verify that in the Hall MHD model, the dispersion relation of the asymptotic symbol $H_{0}$ can accurately predict the topological edge modes admitted by a PT-symmetric non-Hermitian operator $\mathcal{H}$.

\section{Summary}

We studied topological edge modes and spectral flows in non-Hermitian inhomogeneous continuous media that are PT-symmetric and asymptotically Hermitian. For these media, if the asymptotic symbol $H_{0}$ supports a Weyl point of two-fold degeneracy, then the non-Hermitian operator $\mathcal{H}$ of the system admits topological edge modes with real eigenfrequencies, as characterized by a spectral flow near the Weyl point. In other words, with the protection of PT-symmetry, the spectral flow induced by a Weyl point in the Hermitian bulk is robust against non-Hermitian perturbation at the interface in classical continuous media. As an application of the general theory developed, we identified a topological edge mode called topological Alfv\'{e}n-sound wave using the non-Hermitian Hall MHD model. The analysis reported in this paper could be applied to search for topological modes in a wide range of non-Hermitian PT-symmetric inhomogeneous continuous media in fluids and plasmas.

\begin{acknowledgments}
    This research was supported by the US Department of Energy through Contracts No. DE-AC02-09CH1146 (PPPL) and DE-AC52-07NA27344 (LLNL), LLNL-JRNL-863977.
\end{acknowledgments}

\appendix

\section{Spectra of the two-band model with anti-Hermitian parts\label{sec:analytical_spectral_flow}}

In this section, we calculate the spectra and the eigenmodes of the following class of operators, 
\begin{align}
\mathcal{H}[\lambda;x,\partial_{x}] & =\begin{pmatrix}f(x)/\kappa & \lambda-\partial_{x}\\
\lambda+\partial_{x} & -\kappa f(x)
\end{pmatrix}+\mathrm{i}g(x)\sigma^{\alpha}\label{eq:tilted_H}\\
 & \doteq\mathcal{H}_{\mathrm{H}}+\mathrm{i}\mathcal{H}_{\mathrm{A}}.
\end{align}
Here, $f(x)$ and $g(x)$ are real functions of $x$, and $\lambda$ is a constant control parameter. The parameter $\kappa$ measures the tilting of the Dirac operator \cite{qin2022topological}, indicating that $f(x)$ is associated to both $\sigma^{z}$ and $\sigma^{0}$. Equation (\ref{eq:2_by_2_demo}) corresponds to the case where $\kappa=1$ and $g(x)=\epsilon f'(x)$. We first give the analytical solutions for several special cases, and then provide numerically solved examples for general $f$ and $g$.

\subsection{Analytical solution without anti-Hermitian part}

When $f(x)=x$ and $g=0$, the operator only has a Hermitian part, 
\begin{align}
\mathcal{H}_{\mathrm{H}}=\begin{pmatrix}x/\kappa & \lambda-\partial_{x}\\
\lambda+\partial_{x} & -\kappa x
\end{pmatrix},
\end{align}
which can be analytically solved using the techniques given in Refs.~\cite{faure2023manifestation,qin2022topological,delplace2022berry}. Let $R=\mathrm{diag}(\kappa,1)$, we have 
\begin{align}
\mathcal{H}_{\mathrm{H}}'=R^{-1}\mathcal{H}_{\mathrm{H}}R=\begin{pmatrix}{x}/\kappa & (\lambda-\partial_{x})/\kappa\,\\
\kappa(\lambda+\partial_{x}) & -\kappa{x}
\end{pmatrix}
\end{align}
Define the following variables for convenience, 
\begin{align}
\mu_{1}\doteq\dfrac{1}{2}\left(\kappa+\dfrac{1}{\kappa}\right),\quad\mu_{2}\doteq\dfrac{1}{2}\left(\kappa-\dfrac{1}{\kappa}\right).
\end{align}
In terms of Pauli matrices, 
\begin{align}
    \begin{split}
        \mathcal{H}_{\mathrm{H}}'
        = & -\mu_{2}x\sigma^{0}
        +(\mu_{2}\partial_{x}+\mu_{1}\lambda)\sigma^{x} \\
        & -\mathrm{i}(\mu_{1}\partial_{x}+\mu_{2}\lambda)\sigma^{y}
        +\mu_{1}x\sigma^{z}.
    \end{split}
\end{align}
Making a cyclic rotation of Pauli matrices $(\sigma^{0},\sigma^{x},\sigma^{y},\sigma^{z})\to(\sigma^{0},\sigma^{z},\sigma^{x},\sigma^{y})$ through a similarity transformation, $\mathcal{H}_{\mathrm{H}}'$ is transformed to 
\begin{align}
    \begin{split}
        \mathcal{H}_{\mathrm{H}}''
        = & -\mu_{2}x\sigma^{0}+(\mu_{2}\partial_{x}+\mu_{1}\lambda)\sigma^{z} \\ 
        & -\mathrm{i}(\mu_{1}\partial_{x}+\mu_{2}\lambda)\sigma^{x}+\mu_{1}x\sigma^{y}.
    \end{split}\\[3pt]
 = & \begin{pmatrix}\mu_{1}\lambda-\sqrt{2}\mu_{2}\hat{a}^{\dagger} & \mathrm{i}(-\mu_{2}\lambda-\sqrt{2}\mu_{1}\hat{a})\\
\mathrm{i}(-\mu_{2}\lambda+\sqrt{2}\mu_{1}\hat{a}^{\dagger}) & -\mu_{1}\lambda-\sqrt{2}\mu_{2}\hat{a}
\end{pmatrix}.
\end{align}
Here, 
\begin{align}
\hat{a}\doteq\frac{1}{\sqrt{2}}\left(x+\partial_{x}\right),\quad\hat{a}^{\dagger}\doteq\frac{1}{\sqrt{2}}\left(x-\partial_{x}\right)
\end{align}
are the ladder operators.

Define functions $|n;\delta\rangle$ as 
\begin{align}
    \begin{split}
        \langle x|n;\delta\rangle &\doteq \varphi_{n}(x+\sqrt{2}\delta),\\ 
        \varphi_{n}(x) &\doteq \frac{1}{\left(2^{n}n!\sqrt{\pi}\right)^{1/2}}\mathrm{e}^{-{x^{2}}/{2}}H_{n}(x),
    \end{split}\label{eq:hermite}
\end{align}
where $H_{n}(x)$ is the $n$-th Hermite polynomial. The eigenvectors of the bulk modes of $\mathcal{H}''_{\mathrm{H}}$ are 
\begin{align}
\Psi_{n}^{\pm}=\begin{pmatrix}|n;\delta_{n}^{\pm}\rangle\\
\mathrm{i}\gamma_{n}^{\pm}|n+1;\delta_{n}^{\pm}\rangle
\end{pmatrix},\quad n=0,1,2,\ldots,
\end{align}
where 
\begin{align}
    \begin{split}
        \gamma_{n}^{\pm} &=\dfrac{-\lambda\pm\sqrt{\lambda^{2}+n+1}}{\sqrt{n+1}},\\ 
        \delta_{n}^{\pm} &=\mp\dfrac{\mu_{2}}{\mu_{1}}\sqrt{\lambda^{2}+n+1}.
    \end{split}
\end{align}
The corresponding eigenvalues are 
\begin{align}
\omega_{n}^{\pm}=\pm\dfrac{2\sqrt{2}\kappa}{1+\kappa^{2}}\sqrt{\lambda^{2}+n+1}.
\end{align}
We also find the eigenvector and eigenvalue of the spectral flow of $\mathcal{H}''_{\mathrm{H}}$, labeled by $n=-1$, to be 
\begin{align}
\Psi_{-1} &= \begin{pmatrix}0\\
|0,\delta_{-1}\rangle
\end{pmatrix},\\ 
\delta_{-1} &= \dfrac{\mu_{2}}{\mu_{1}}\lambda,\quad\omega_{-1}=-\dfrac{2\sqrt{2}\kappa}{1+\kappa^{2}}\lambda.
\end{align}

\subsection{Analytical solutions with anti-Hermitian parts\label{sec:analytical_non_hermitian}}

Here, we assume $f(x)=x,\,g(x)=f'(x)=1$ and study the eigenvalues and eigenvectors of the Hamiltonian operator $\mathcal{H}$ in Eq.\,(\ref{eq:tilted_H}) with different $\mathrm{i}\sigma^{\alpha}$. We will see that when $\kappa\neq1$, only the $\mathcal{H}$ with $\mathrm{i}\sigma^{y}$ renders the system PT-symmetric.

\textit{1.} $\mathcal{H}$ with\textit{ $\mathrm{i}\sigma^{0}$.} Since $\mathrm{i}\sigma^{0}$ does not depend on $x$, the addition of $\mathrm{i}\sigma^{0}$ only shifts the eigenvalue by $\mathrm{i}$:
\begin{align}
\begin{cases}
\omega_{n}^{\pm}=\pm\dfrac{2\sqrt{2}\kappa}{1+\kappa^{2}}\sqrt{\lambda^{2}+n+1}+\mathrm{i},\quad n \geq 0,\\[10pt]
\omega_{-1}=-\dfrac{2\sqrt{2}\kappa}{1+\kappa^{2}}\lambda+\mathrm{i}.
\end{cases}
\end{align}

\textit{2. }$\mathcal{H}$ with\textit{ $\mathrm{i}\sigma^{x}$.} In terms of slightly modified control parameter $\tilde{\lambda}\doteq\lambda+\mathrm{i}$, 
\begin{align}
\mathcal{H}=\begin{pmatrix}x/\kappa & \tilde{\lambda}-\partial_{x}\\
\tilde{\lambda}+\partial_{x} & -\kappa x
\end{pmatrix}.
\end{align}
Therefore, the eigenvalues are 
\begin{align}
\begin{cases}
\omega_{n}^{\pm}=\pm\dfrac{2\sqrt{2}\kappa}{1+\kappa^{2}}\sqrt{(\lambda+\mathrm{i})^{2}+n+1},\quad n \geq 0,\\[10pt]
\omega_{-1}=-\dfrac{2\sqrt{2}\kappa}{1+\kappa^{2}}(\lambda+\mathrm{i}).
\end{cases}
\end{align}

\textit{3. }$\mathcal{H}$ with\textit{ $\mathrm{i}\sigma^{y}$.} By applying the technique described in the main text, the anti-Hermitian part $\mathrm{i}\sigma^{y}$ in operator $\mathcal{H}$ can be removed by a similarity transformation $\mathcal{H\rightarrow}e^{-x}\mathcal{H}e^{x}$. The eigenvalues thus do not change after adding $\mathrm{i}\sigma^{y}$. Noticeably, the factor $e^{-x}$ in the eigenvectors resembles the ``non-Hermitian skin effect'' \cite{yao2018edge,yao2018non} found in general non-Hermitian systems. However, in the given example, eigenvectors are already localized at the interface due to the factor $e^{-x^{2}/2}$ in Eq.~(\ref{eq:hermite}), even in the Hermitian case.

\textit{4. }$\mathcal{H}$ with\textit{ $\mathrm{i}\sigma^{z}$.} In terms of modified variable $\tilde{x}\doteq x+\mathrm{i}/\mu_{1}$, the operator is 
\begin{align}
\mathcal{H}=\begin{pmatrix}\tilde{x}/\kappa & \lambda-\partial_{\tilde{x}}\\
\lambda+\partial_{\tilde{x}} & -\kappa\tilde{x}
\end{pmatrix}+\dfrac{\mu_{2}}{\mu_{1}}\mathrm{i}\sigma^{0}.
\end{align}
The eigenvalues are shifted by $\mathrm{i}\mu_{2}/\mu_{1}$, similar to the case of $\mathcal{H}$ with $\mathrm{i}\sigma^{0}$. When $\kappa=1$, the operator is not tilted, and the addition of $\mathrm{i}\sigma^{z}$ does not introduce an imaginary part.

\subsection{Numerical solutions with anti-Hermitian parts}

\begin{figure}[ht]
\centering 
\centering 
\includegraphics[height=4cm]{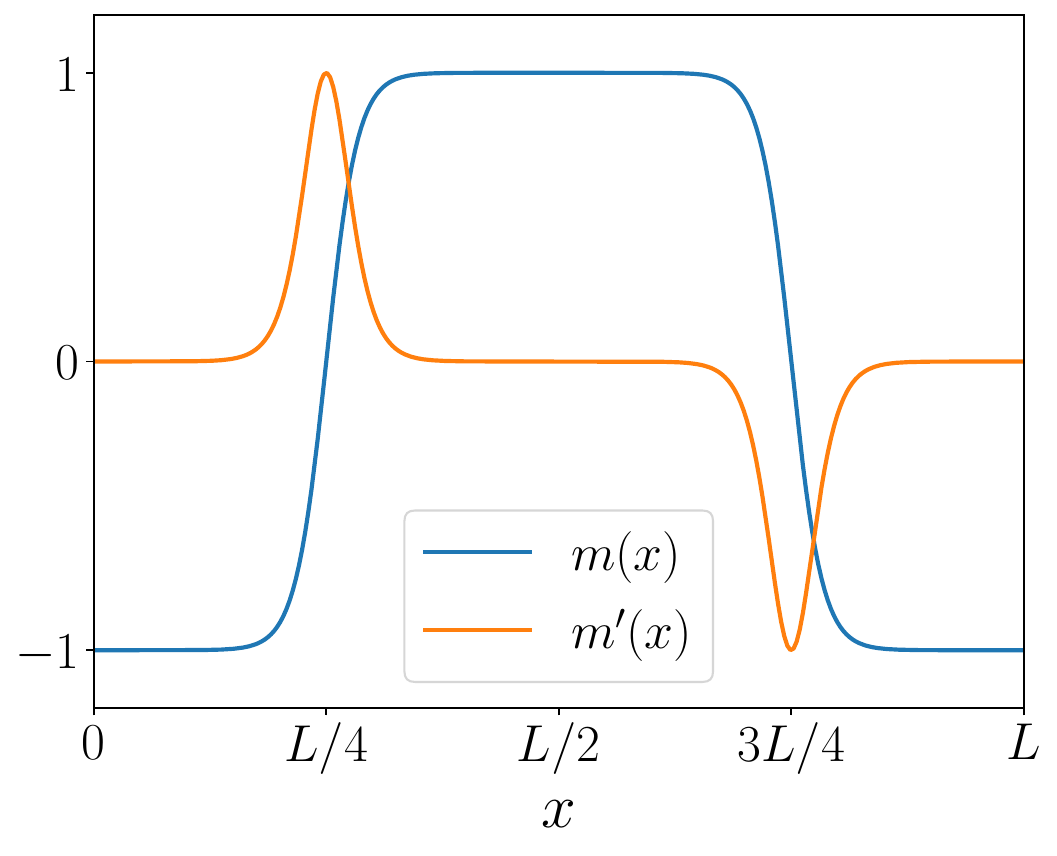} 
\caption{$m(x)$ and $m'(x)$.}
\label{fig:m_profile}
\end{figure}

Here we present several numerical results on the spectra of $\mathcal{H}$ in Eq.~(\ref{eq:tilted_H}) for the case of $f(x)=m(x)$, $g(x)=\epsilon m'(x)$, and 
\begin{align}
m(x)\doteq\tanh\left(\dfrac{x-\frac{1}{4}L}{l}\right)-\tanh\left(\dfrac{x-\frac{3}{4}L}{l}\right)-1.
\end{align}
Here, $L$ is the system length, $l$ is the scale length of the interface region. $m(x)$ and $m'(x)$ are plotted in Fig.~\ref{fig:m_profile}. The $x$-direction is chosen to be periodic with periodicity $L$. In the bulk region around $x=L/2$ and $x=0$, $m'=0$ and the system is homogeneous. In the interface region at $x=L/4$ and $x=3L/4$, $m'(x)\neq0$ and system is non-Hermitian.

\begin{figure}[ht]
\centering 
\includegraphics[width=0.55\columnwidth]{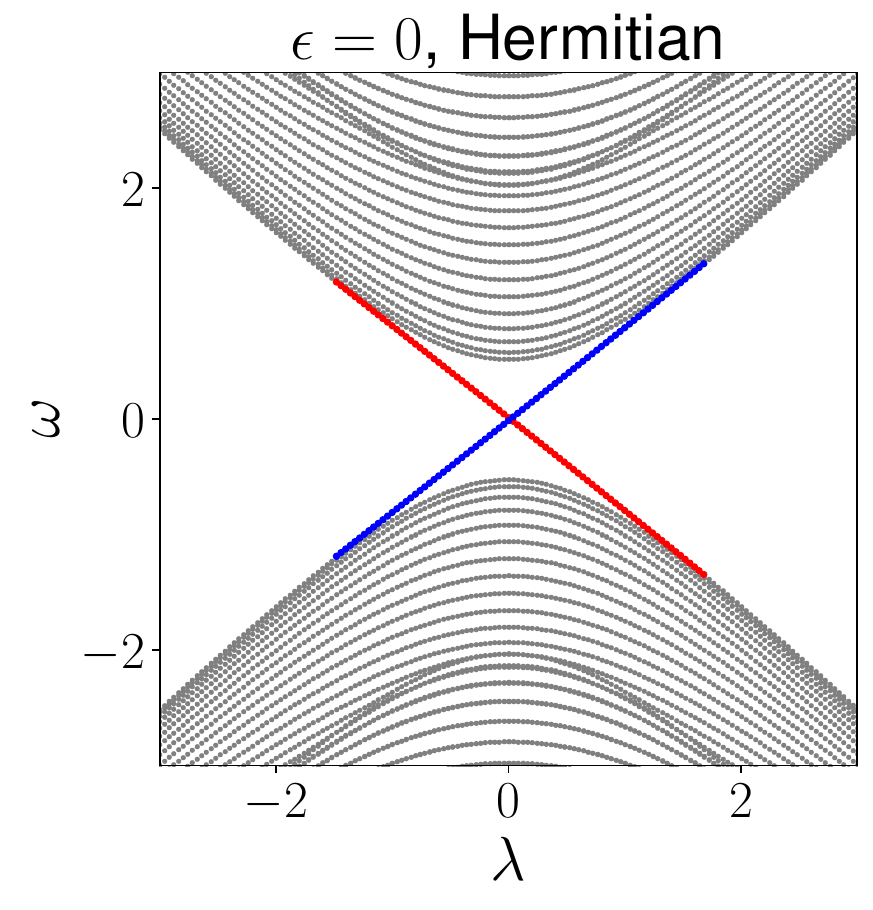} 
\caption{The spectrum of Hermitian operators. Grey dots indicate bulk mods, red/blue dots indicate topological edge modes (spectral flow) localized at the left/right interface.}
\label{fig:m_profile_hermitian}
\end{figure}

In the numerical example, we choose $L=30,l=1,$ and $\kappa=2$. The spectrum of the Hermitian operator at $\epsilon=0$ is shown in Fig.~\ref{fig:m_profile_hermitian}. Since there are two interface regions at $x=L/4$ and $x=3L/4$, there are two spectral flows indicated by the blue and red dots.

\begin{figure}[ht]
\centering 
\includegraphics[width=0.95\columnwidth]{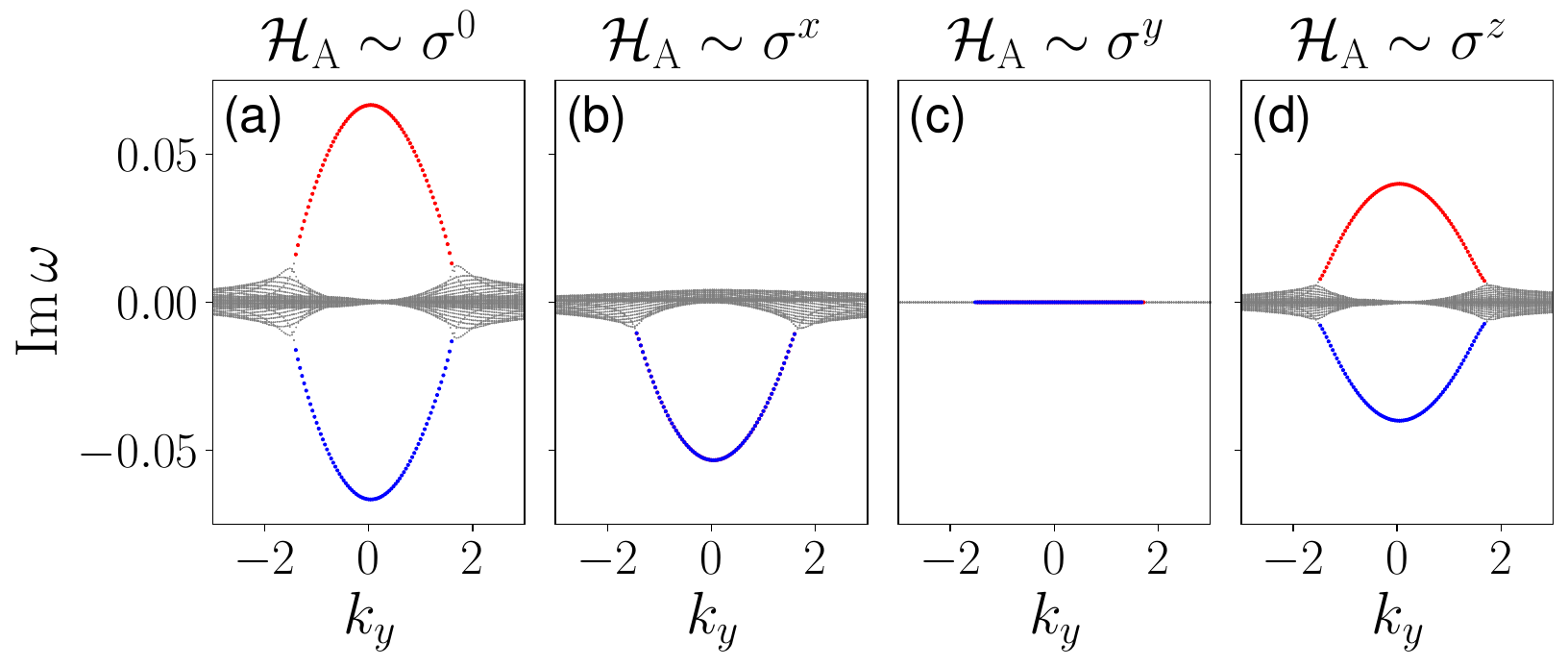} 
\caption{The imaginary part of operator $\mathcal{H}$ in Eq.\,(\ref{eq:tilted_H}), whose anti-Hermitian part $\mathcal{H}_{\mathrm{A}}$ is proportional to (a) $\sigma^{0}$, (b) $\sigma^{x}$, (c) $\sigma^{y}$, and (d) $\sigma^{z}$. The gray points indicate bulk modes, the blue/red points indicate edge modes localized at left/right edge. In (b) and (c), the blue and red points coincide.}
\label{fig:m_profile_antihermitian}
\end{figure}

For the non-Hermitian case, we choose $\epsilon=0.1$, so that the real part of the spectrum of $\mathcal{H}$ does not deviate significantly from its Hermitian counterpart. With different anti-Hermitian parts $\mathrm{i}\sigma^{\alpha}$, the real parts of the spectrum are qualitatively similar to Fig.~\ref{fig:m_profile_hermitian}. The imaginary parts are shown in Fig.~\ref{fig:m_profile_antihermitian}. 

We observe that with anti-Hermitian terms proportional to $\mathrm{i}\sigma^{0},\,\mathrm{i}\sigma^{x}$ or $\mathrm{i}\sigma^{z}$, the spectrum of $\mathcal{H}$, including its spectral flow, are complex. The effect of $\mathrm{i}\sigma^{0}$ and $\mathrm{i}\sigma^{z}$ are similar, akin to the analytical solutions given above. However, with an anti-Hermitian term proportional to $\mathrm{i}\sigma^{y}$, $\mathcal{H}$ is PT-symmetric, and its spectrum is still real.

\section{The Hall MHD model\label{sec:hmhd_equations}}

Within the MHD model, which focuses on low frequency and long-wavelength phenomena, the displacement current $\partial_{t}\mathbf{E}$ in Maxwell's equations is ignored. When studying plasma waves, the dynamics of plasma is usually assumed to be adiabatic with the equation of state $p/\rho^{\gamma}=\mathrm{Const.}$, where $p$ is plasma pressure, $\rho$ is plasma mass density, and $\gamma$ is ratio of specific heats. With this assumption, the equations of motions for $\rho,$ $\mathbf{v}$, $\mathbf{B}$, and $p$ are 
\begin{align}
 & \partial_{t}\rho+\nabla\cdot(\rho\mathbf{v})=0,\label{eq:mass}\\[3pt]
\text{} & \rho\dfrac{\mathrm{d}\mathbf{v}}{\mathrm{d}t}=\mathbf{j}\times\mathbf{B}-\nabla p,\label{eq:momentum}\\[3pt]
\text{} & \partial_{t}\mathbf{B}=-\nabla\times\mathbf{E},\label{eq:induction}\\[3pt]
 & \dfrac{\mathrm{d}p}{\mathrm{d}t}=-\gamma p\nabla\cdot\mathbf{v},\label{eq:state}
\end{align}
where $\mathbf{v}$ is the plasma macroscopic velocity, and $\mathrm{d}/\mathrm{d}t\doteq\partial_{t}+\mathbf{v}\cdot\nabla$. The plasma current $\mathbf{j}$ is related to the magnetic field $\mathbf{B}$ by the Ampere's law 
\begin{align}
 & \quad\mu_{0}\mathbf{j}=\nabla\times\mathbf{B},
\end{align}
where $\mu_{0}$ is the vacuum permeability. Since the displacement current is ignored, $\mathbf{E}$ is obtained from the generalized Ohm's law. Different forms of Ohm's laws lead to different MHD models. In the Hall MHD approximation, the electron inertia is ignored ($m_{\mathrm{e}}\to0$) and the plasma is assumed to be infinitely conductive. The Ohm's law could be written as \cite{hameiri2004linear} 
\begin{align}
 & \mathbf{E}+\mathbf{v}\times\mathbf{B}=\dfrac{\mathbf{j}\times\mathbf{B}-\nabla p_{\mathrm{e}}}{n_{\mathrm{e}}e},\label{eq:ohm_hall}
\end{align}
where $e>0$ is the elementary charge, $n_{\mathrm{e}}$ is the electron number density, $p_{\mathrm{e}}$ is the electron pressure.

Combining Faraday's law and Ohm's law, we can eliminate electric field $\mathbf{E}$ 
\begin{align}
\begin{split}
    \partial_{t}\mathbf{B}
    = & \nabla\times(\mathbf{v}\times\mathbf{B}) \\
    & -\nabla\times\left(\dfrac{\mathbf{j}\times\mathbf{B}}{n_{\mathrm{e}}e}\right)+\nabla\times\left(\dfrac{\nabla p_{\mathrm{e}}}{n_{\mathrm{e}}e}\right).
\end{split}
\label{eq:eliminate_E}
\end{align}
If the electrons are assumed to be barotropic \cite{hameiri2004linear}, i.e., the electron pressure is a function of electron number density, $p_{\mathrm{e}}=p_{\mathrm{e}}(n_{\mathrm{e}})$, the last term in Eq.~(\ref{eq:eliminate_E}) vanishes.

Consider an equilibrium without flow and with homogeneous densities $\rho$ and $n_{\mathrm{e}}$. Hereafter, we use symbols without subscripts such $\mathbf{B}$, $p$, and $\mathbf{j}$ to indicate equilibrium, time-independent fields. From Eq.\,(\ref{eq:momentum}), the equilibrium fields $\mathbf{B}$ and $p$ satisfy the pressure balance, 
\begin{align}
\nabla p=\mathbf{j}\times\mathbf{B}=\dfrac{1}{\mu_{0}}(\nabla\times\mathbf{B})\times\mathbf{B}.
\end{align}

For a given equilibrium, the perturbed field $\left(\rho_{1},\mathbf{v}_{1},\mathbf{B}_{1},\mathbf{p}_{1}\right)$ are governed by the linearized system, 
\begin{align}
\partial_{t}\rho_{1} =& -\rho\nabla\cdot\mathbf{v}_{1},\label{eq:linearized_mass}\\[3pt]
\rho\partial_{t}\mathbf{v}_{1} =& \mathbf{j}_{1}\times\mathbf{B}+\mathbf{j}\times\mathbf{B}_{1}-\nabla p_{1},\\[3pt]
\begin{split}
    \partial_{t}\mathbf{B}_{1} =&  \nabla\times(\mathbf{v}_{1}\times\mathbf{B}) \\
    & -\dfrac{1}{en_{e}}\nabla\times(\,\mathbf{j}_{1}\times\mathbf{B}+\mathbf{j}\times\mathbf{B}_{1}),
\end{split}
\\[3pt]
\partial_{t}p_{1} =& -\mathbf{v}_{1}\cdot\nabla p-\gamma p\nabla\cdot\mathbf{v}_{1}.
\end{align}
Equation (\ref{eq:linearized_mass}) is decoupled from the others. In terms of the following normalized variables 
\begin{align}
\tilde{\mathbf{v}}\doteq\sqrt{\rho}\mathbf{v}_{1},\quad\tilde{\mathbf{B}}\doteq\dfrac{\mathbf{B}_{1}}{\sqrt{\mu_{0}}},\quad\tilde{p}\doteq\dfrac{p_{1}}{\sqrt{\gamma p}},
\end{align}
the linearized equations become are 
\begin{align}
\partial_{t}\tilde{\mathbf{v}} =& (\nabla\times\tilde{\mathbf{B}})\times\mathbf{v}_{\mathrm{A}}+(\nabla\times\mathbf{v}_{\mathrm{A}})\times\tilde{\mathbf{B}}-\nabla(v_{\mathrm{s}}\tilde{p}),\\[5pt]
\begin{split}
    \partial_{t}\tilde{\mathbf{B}} =& \nabla{\times}(\tilde{\mathbf{v}}\times\mathbf{v}_{\mathrm{A}}) \\
    & -\sqrt{\dfrac{\rho}{\mu_{0}}}\dfrac{1}{n_{\mathrm{e}}e}\nabla{\times}
    \left[(\nabla{\times}\tilde{\mathbf{B}}){\times}\mathbf{v}_{\mathrm{A}}\right] \\ 
    & -\sqrt{\dfrac{\rho}{\mu_{0}}}\dfrac{1}{n_{\mathrm{e}}e}\nabla{\times}
    \left[(\nabla{\times}\mathbf{v}_{\mathrm{A}}){\times}\tilde{\mathbf{B}}\right]
\end{split}
\label{eq:liearized_faraday2}\\[5pt]
\partial_{t}\tilde{p} =& -v_{\mathrm{s}}\nabla\cdot\tilde{\mathbf{v}}-(2/\gamma)\tilde{\mathbf{v}}\cdot\nabla v_{\mathrm{s}},
\end{align}
where 
\begin{align}
\mathbf{v}_{\mathrm{A}}\doteq\dfrac{\mathbf{B}}{\sqrt{\mu_{0}\rho}},\quad v_{\mathrm{s}}\doteq\sqrt{\dfrac{\gamma p}{\rho}}
\end{align}
are Alfv\'{e}n velocity and sound velocity. Let $n_{\mathrm{i}},m_{\mathrm{i}},q_{\mathrm{i}}$ denote the number density, mass, and charge of ions. Using the quasi-neutrality condition $n_{\mathrm{e}}e=n_{\mathrm{i}}q_{\mathrm{i}}$, the coefficient $\sqrt{\rho\mu_{0}/n_{\mathrm{e}}e}$ in Eq.~(\ref{eq:liearized_faraday2}) can be simplified as 
\begin{align}
\begin{split}
    \sqrt{\dfrac{\rho}{\mu_{0}}}\dfrac{1}{n_{\mathrm{e}}e}
    =& \sqrt{\dfrac{n_{\mathrm{i}}m_{\mathrm{i}}}{1/(\varepsilon_{0}c^{2})}}\dfrac{1}{n_{\mathrm{i}}q_{\mathrm{i}}} \\ 
    =& c\left/\sqrt{\dfrac{n_{\mathrm{i}}q_{\mathrm{i}}^{2}}{\varepsilon_{0}m_{\mathrm{i}}}}\right.
    =\dfrac{c}{\omega_{\mathrm{pi}}}
    =d_{\mathrm{i}},
\end{split}
\end{align}
where $\epsilon_{0}$ is the vacuum permittivity, $c$ is the light speed, $\omega_{\mathrm{pi}}$ is the ion plasma frequency, and $d_{\mathrm{i}}$ is known as the ion skin depth. The final linearized equations for Hall MHD become Eqs.~(\ref{eq:linearized_momentum})-(\ref{eq:linearized_state}) in the main text. In the limit of $d_{\mathrm{i}}\to0$, the system reduces to the ideal MHD model where the right-hand of Eq.~(\ref{eq:ohm_hall}) vanishes.

To study the dispersion relation and wave topology in the bulk region, we assume the equilibrium magnetic field $\mathbf{B}$ and pressure $p$ are constants. In this case, the wave fields $\Psi=(\tilde{\mathbf{v}},\tilde{\mathbf{B}},\tilde{p})^{\intercal}$ are governed by $\mathrm{i}\partial_{t}\Psi=\mathcal{H}_{0}\Psi$, where $\mathcal{H}_{0}$ is the asymptotic Hamiltonian shown in Eq.~(\ref{eq:operator_H0}). The corresponding symbol is 
\begin{align}
{H}_{0}=\begin{pmatrix}0 & (\mathbf{v}_{\mathrm{A}}\mathbf{k})^{\intercal}-\mathbf{v}_{\mathrm{A}}{\cdot}\mathbf{k}\,\, & v_{\mathrm{s}}\mathbf{k}\\[3pt]
(\mathbf{v}_{\mathrm{A}}\mathbf{k})-\mathbf{v}_{\mathrm{A}}{\cdot}\mathbf{k}\,\, & \mathrm{i}d_{\mathrm{i}}(\mathbf{v}_{\mathrm{A}}{\cdot}\mathbf{k})\mathbf{k}\times & 0\\[3pt]
v_{\mathrm{s}}\mathbf{k}^{\intercal} & 0 & 0
\end{pmatrix}.\label{eq:symbol_H0}
\end{align}
It is clear that $H_{0}$ and $\mathcal{H}_{0}$ are Hermitian.

\begin{figure}[ht]
\centering \includegraphics[width=0.95\columnwidth]{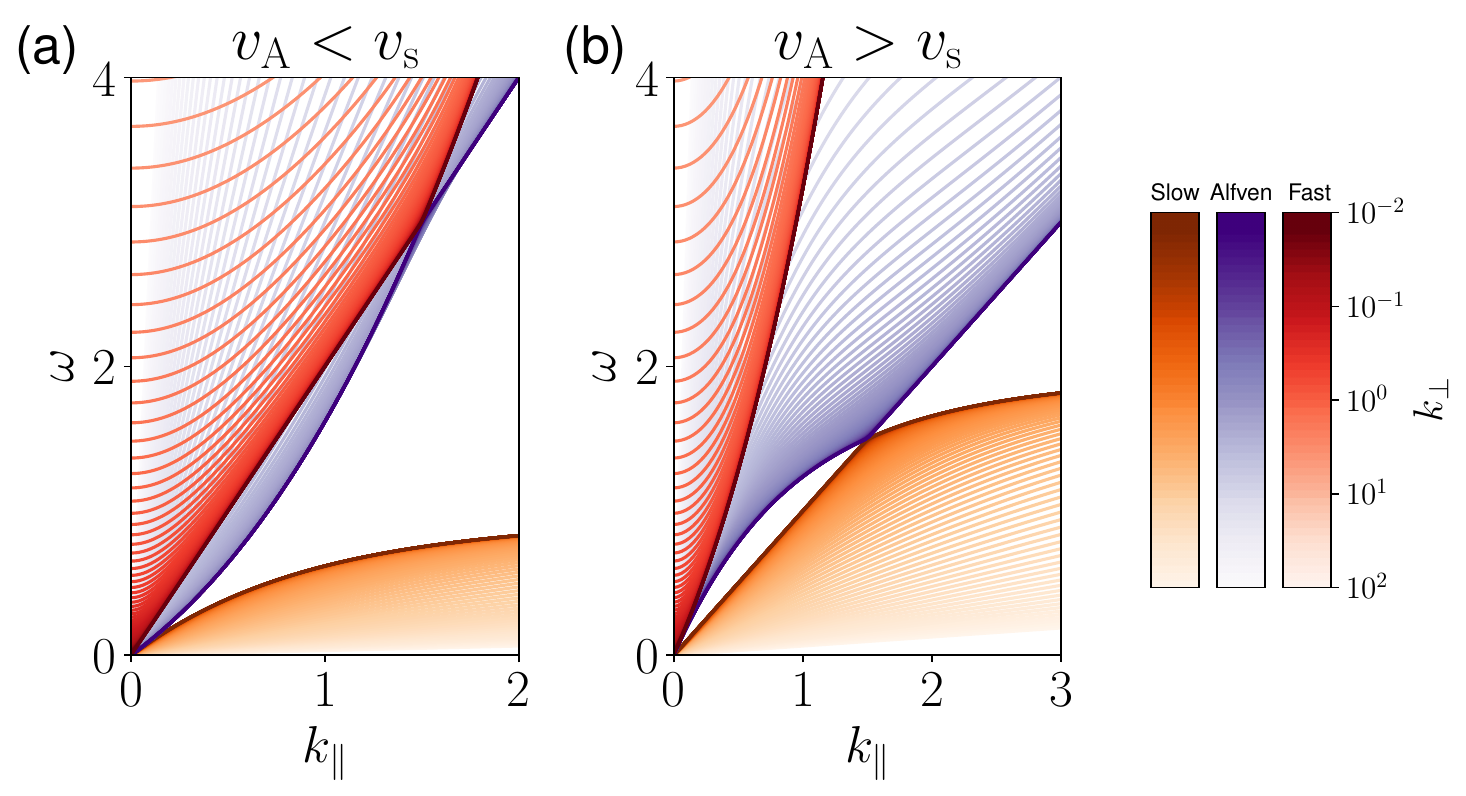} \caption{Dispersion relations of three positive frequency branches. The red, purple and orange lines represent the fast, Alfv\'{e}n, and slow waves, respectively. The saturation level of each color represents the value of $k_{\perp}$. There are two types of Weyl points depending on the ratio between $v_{\mathrm{s}}$ and $v_{\mathrm{A}}$.}
\label{fig:dispersion_surface_2D}
\end{figure}

\begin{figure}[ht]
\centering \includegraphics[width=0.95\columnwidth]{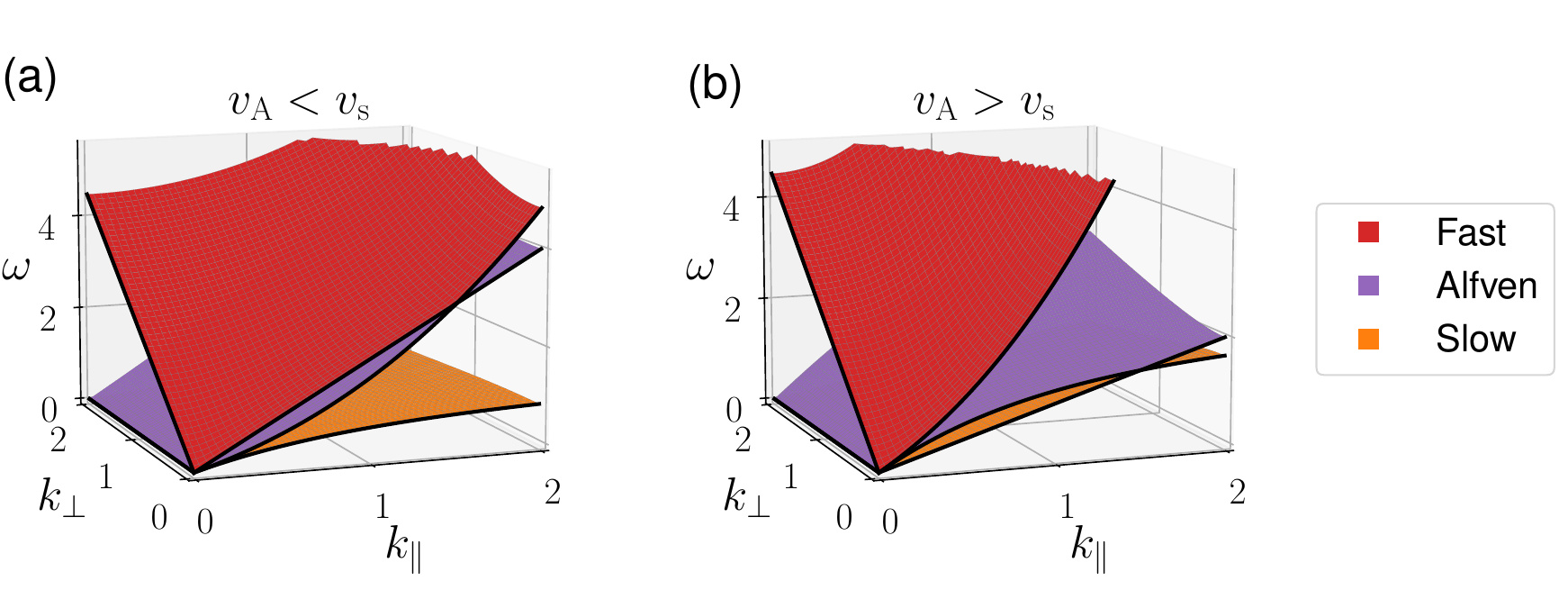} \caption{The 3D rendition of the dispersion relations in Fig.\,\ref{fig:dispersion_surface_2D}.}
\label{fig:dispersion_surface_3D}
\end{figure}

Let magnetic field is in the $z$-direction and $\mathbf{v}_{\mathrm{A}}=v_{\mathrm{A}}\mathbf{e}_{z}$. $H_{0}$ has one zero eigenvalue $\omega_{0}=0$, whose eigenvector is $\Psi=(0,0,0,0,0,1,0)^{\intercal}$. The rest 6 eigenvalues are given by Eq.~(\ref{eq:bulk_dispersion}). Because $\omega$ only appears in terms of $\omega^{2}$ in Eq.~(\ref{eq:bulk_dispersion}), the eigenvalues are symmetric with respect to zero. We can thus only study the three positive frequency branches. They are called fast wave, Alfv\'{e}n wave, and slow wave according to their phase velocities, which satisfies the following order, 
\begin{align}
v_{\mathrm{p,slow}}\leq v_{\mathrm{p,Alfven}}\leq v_{\mathrm{p,fast}}.\label{eq:vpslow}
\end{align}
The dispersion relation is plotted in Figs.~\ref{fig:dispersion_surface_2D} and \ref{fig:dispersion_surface_3D}. Since $k_{x}$ and $k_{y}$ contribute symmetrically through $k_{\perp}^{2}\doteq k_{x}^{2}+k_{y}^{2}$, the eigenvalues $\omega_{n}$ are plotted as functions of $(k_{\perp},k_{z})$. 

From the dispersion relations, we observe that a Weyl point can be formed by the crossing between the fast and Alfv\'{e}n waves or between the slow and Alfv\'{e}n waves. The Weyl point can be analytically solved for from Eq.~(\ref{eq:bulk_dispersion}). When $k_{\perp}=0$, Eq.~(\ref{eq:bulk_dispersion}) has three positive solutions, 
\begin{align}
\omega_{1}&=v_{s}k_{z},\\
\omega_{2,3}&=v_{\mathrm{A}}k_{z}\left(\sqrt{1+\dfrac{1}{4}k_{z}^{2}d_{\mathrm{i}}^{2}}\pm\dfrac{1}{2}k_{z}d_{\mathrm{i}}\right).
\end{align}
The value of $k_{z}$ at the Weyl points can be expressed in terms of $\zeta=v_{\mathrm{s}}/v_{\mathrm{A}}$, 
\begin{align}
k_{z}d_{\mathrm{i}}=\left|\dfrac{1}{\zeta}-\zeta\right|=\begin{cases}
{1}/{\zeta}-\zeta,\quad\text{when: }\zeta<1,\\[3pt]
\zeta-{1}/{\zeta},\quad\text{when: }\zeta>1.
\end{cases}
\end{align}
If $k_{z}$ is fixed, the value of $\xi$ at the Weyl points is 
\begin{align}
\zeta=\sqrt{1+\dfrac{1}{4}k_{z}^{2}d_{\mathrm{i}}^{2}}\pm\dfrac{1}{2}k_{z}d_{\mathrm{i}}.
\end{align}

For a fixed $k_{z}$, the topological charge of each Weyl point can be calculated in the parameter space $(x,k_{x},k_{y})$, which is given explicitly using a two-band approximation in Appendix~\ref{sec:TAS}.

\section{The topological Alfv\'{e}n-sound wave in Hall MHD \label{sec:TAS}}

In this section, we provide detailed derivations of the topological Alfv\'{e}n-sound wave in Hall MHD.

In an inhomogeneous equilibrium, the operator $\mathcal{H}$ is relatively complicated. From Eqs.~(\ref{eq:linearized_momentum})-(\ref{eq:linearized_state}), the operator $\mathcal{H}$ can be written in a $7\times7$ matrix, 
\begin{align}
\mathcal{H} & =\mathcal{H}_{0}+\B{\mathcal{H}_{1}}+\R{\mathcal{H}_{2}}
=  \left(\begin{array}{ccc}
0 & h_{12} & h_{13}\\
h_{21} & h_{22} & 0\\
h_{31} & 0 & 0
\end{array}\right),
\label{eq:hmhd_operator}
\end{align}
where

\begin{align*}
h_{12} & =\left(\begin{array}{ccc}
-v_{\mathrm{A}}k_{z} & 0 & -\mathrm{i}v_{\mathrm{A}}\partial_{x}\B{-\mathrm{i}v_{\mathrm{A}}'}\\
0 & -v_{\mathrm{A}}k_{z} & v_{\mathrm{A}}k_{y}\\
\B{\mathrm{i}v_{\mathrm{A}}'} & 0 & 0
\end{array}\right),\\[3pt]
h_{13} & =\left(\begin{array}{c}
-\mathrm{i}v_{\mathrm{s}}\partial_{x}\B{-\mathrm{i}v_{\mathrm{s}}'}\\
v_{\mathrm{s}}k_{y}\\
v_{\mathrm{s}}k_{z}
\end{array}\right),\\[3pt]
h_{21} & =
\left(\begin{array}{ccc}
-v_{\mathrm{A}}k_{z} & 0 & 0\\
0 & -v_{\mathrm{A}}k_{z} & 0\\
-\mathrm{i}v_{\mathrm{A}}\partial_{x}\B{-\mathrm{i}v_{\mathrm{A}}'} & v_{\mathrm{A}}k_{y} & 0
\end{array}\right),\\[3pt]
h_{22} & =
\left(\begin{array}{ccc}
\B{d_{\mathrm{i}}v_{\mathrm{A}}'k_{y}} & -\mathrm{i}d_{\mathrm{i}}v_{\mathrm{A}}k_{z}^{2} & \mathrm{i}d_{\mathrm{i}}v_{\mathrm{A}}k_{y}k_{z}\\
\mathrm{i}d_{\mathrm{i}} (v_{\mathrm{A}}k_{z}^{2}\R{+v_{\mathrm{A}}''}) & \B{d_{\mathrm{i}}v_{\mathrm{A}}'k_{y}} & -k_{z}d_{\mathrm{i}}v_{\mathrm{A}}\partial_{x}\\
-\mathrm{i}d_{\mathrm{i}}v_{\mathrm{A}}k_{y}k_{z} & d_{\mathrm{i}}k_{z}(v_{\mathrm{A}}\partial_{x}\B{+v_{\mathrm{A}}'}) & 0
\end{array}\right),\\[3pt]
h_{31} & =\left(\begin{array}{ccc}
-\mathrm{i}v_{\mathrm{s}}\partial_{x}\B{-\mathrm{i}(2/\gamma)v_{\mathrm{s}}'} & v_{\mathrm{s}}k_{y} & v_{\mathrm{s}}k_{z}\end{array}\right).
\end{align*}
Here, $\mathcal{H}_{0}$ is the asymptotic Hamiltonian that depends on $v_{\mathrm{A}}$ and $v_{\mathrm{s}}$, but not their derivatives. It is the same as Eq.~(\ref{eq:operator_H0}) in the main text. The terms that show up in $\mathcal{H}_{0}$ are written in black. We can see that $\mathcal{H}$ is significantly different from $\mathcal{H}_{0}$ in the inhomogeneous region. $\mathcal{H}_{1}$, which are highlighted in blue, depends on the first order derivatives. Similarly, $\mathcal{H}_{2}$, highlighted in red, depends on the second order derivatives. It is clear that $\mathcal{H}$ is not Hermitian due to the background inhomogeneity. The anti-Hermitian part of $\mathcal{H}$ is 
\begin{widetext}
\begin{align}
\mathcal{H}_{\mathrm{A}}\doteq\dfrac{1}{2\mathrm{i}}(\mathcal{H}-\mathcal{H}^{\dagger})=\left(\begin{array}{ccc|ccc|c}
0 & 0 & 0 & 0 & 0 & -v_{\mathrm{A}}'/2 & -v_{\mathrm{s}}'/\gamma\\
0 & 0 & 0 & 0 & 0 & 0 & 0\\
0 & 0 & 0 & v_{\mathrm{A}}'/2 & 0 & 0 & 0\\
\hline 0 & 0 & v_{\mathrm{A}}'/2 & 0 & {d}_{\mathrm{i}}v_{\mathrm{A}}''/2 & 0 & 0\\
0 & 0 & 0 & {d}_{\mathrm{i}}v_{\mathrm{A}}''/2 & 0 & 0 & 0\\
-v_{\mathrm{A}}'/2 & 0 & 0 & 0 & 0 & 0 & 0\\
\hline -v_{\mathrm{s}}'/\gamma & 0 & 0 & 0 & 0 & 0 & 0
\end{array}\right)
\end{align}
\end{widetext}
Notice that for a real function $f(x)$, 
\begin{align}
\left[\mathrm{i}f(x)\dfrac{\mathrm{d}}{\mathrm{d}x}\right]^{\dagger}=\mathrm{i}f'(x)+\mathrm{i}f(x)\dfrac{\mathrm{d}}{\mathrm{d}x}.
\end{align}

\begin{figure}[ht]
\centering 
\includegraphics[width=0.95\columnwidth]{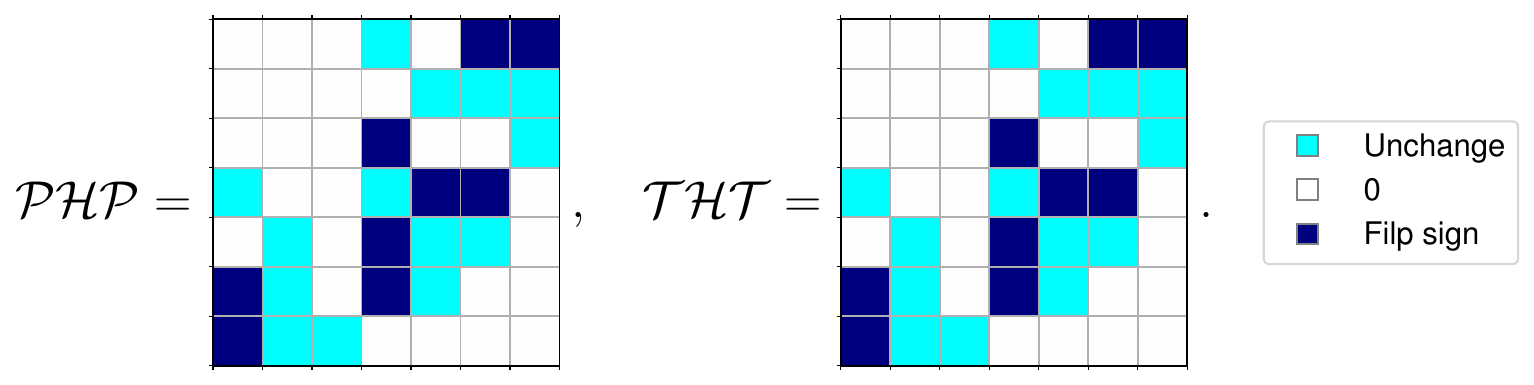} 
\caption{The effects of $\mathcal{P}$ and $\mathcal{T}$ operators on $\mathcal{H}$. The elements of $\mathcal{H}$ are represented by a $7\times7$ grids. Vanishing elements are shown in white. Elements that are unchanged under an operator are shown in light blue. Elements that have their sign flipped under an operator are shown in dark blue.}
\label{fig:PT_symmetry}
\end{figure}

Although $\mathcal{H}$ is not Hermitian, it is PT-symmetric for $\mathcal{P}=\mathrm{diag}(-1,1,1,-1,1,1,1)$ and $\mathcal{T}$ being complex conjugation, i.e., $\mathcal{PT}\mathcal{H}\mathcal{TP}=\mathcal{P}\mathcal{H}^{*}\mathcal{P}=\mathcal{H}$. To prove this property, the effect of $\mathcal{P}$ and $\mathcal{T}$ operators on $\mathcal{H}$ are graphically shown in Fig.~\ref{fig:PT_symmetry}. Each element in Eq.~(\ref{eq:hmhd_operator}) is either real or imaginary. Under time reversal or parity transform, the sign of each element is either unchanged or flipped. Observe that $\mathcal{P}$ and $\mathcal{T}$ have the same effect on $\mathcal{H}$. Thus, their combination keeps $\mathcal{H}$ unchanged. 

For a function $f(x)$ and operator $\hat{k}=-\mathrm{i}\partial_{x}$, their Wigner-Weyl transforms are 
\begin{align}
&\mathscr{W}[f(x)]=f(x),
\quad\mathscr{W}[\hat{k}]=k, \\
&\mathscr{W}[f(x)\hat{k}]=f(x)k+\mathrm{i}f'(x)/2.
\end{align}
From these relations, the symbol of $\mathcal{H}$ in Eq.~(\ref{eq:hmhd_operator}) is 
\begin{align*}
H & =\mathscr{W}[\mathcal{H}]=H_{0}+\B{H_{1}}+\R{H_{2}}
=\left(\begin{array}{ccc}
0 & s_{12} & s_{13}\\
s_{21} & s_{22} & 0\\
s_{31} & 0 & 0
\end{array}\right),
\end{align*}
where 
\begin{align*}
s_{12} & =\left(\begin{array}{ccc}
-v_{\mathrm{A}}k_{z} & 0 & v_{\mathrm{A}}k_{x}\B{-\frac{\mathrm{i}}{2}v_{\mathrm{A}}'}\\
0 & -v_{\mathrm{A}}k_{z} & v_{\mathrm{A}}k_{y}\\
\B{\mathrm{i}v_{\mathrm{A}}'} & 0 & 0
\end{array}\right),\\
s_{13} & =\left(\begin{array}{c}
v_{\mathrm{s}}k_{x}\B{-\frac{\mathrm{i}}{2}v_{\mathrm{s}}'}\\
v_{\mathrm{s}}k_{y}\\
v_{\mathrm{s}}k_{z}
\end{array}\right),\\
s_{21} & =\left(\begin{array}{ccc}
-v_{\mathrm{A}}k_{z} & 0 & 0\\
0 & -v_{\mathrm{A}}k_{z} & 0\\
v_{\mathrm{A}}k_{x}\B{-\frac{\mathrm{i}}{2}v_{\mathrm{A}}'} & v_{\mathrm{A}}k_{y} & 0
\end{array}\right),\\
s_{22} & =\left(\begin{array}{ccc}
\B{d_{\mathrm{i}}v_{\mathrm{A}}'k_{y}} & -\mathrm{i}d_{\mathrm{i}}v_{\mathrm{A}}k_{z}^{2} & \mathrm{i}d_{\mathrm{i}}v_{\mathrm{A}}k_{y}k_{z}\\
\mathrm{i}d_{\mathrm{i}}(v_{\mathrm{A}}k_{z}^{2}\R{+v_{\mathrm{A}}''}) & \B{d_{\mathrm{i}}v_{\mathrm{A}}'k_{y}} & -k_{z}d_{\mathrm{i}}v_{\mathrm{A}}\partial_{x}\\
-\mathrm{i}d_{\mathrm{i}}v_{\mathrm{A}}k_{y}k_{z} & d_{\mathrm{i}}k_{z}(\mathrm{i}v_{\mathrm{A}}k_{x}\B{+\frac{v_{\mathrm{A}}'}{2}}) & 0
\end{array}\right),\\
s_{31} & =\left(\begin{array}{ccc}
v_{\mathrm{s}}k_{x}\B{+\mathrm{i}(\frac{1}{2}-\frac{2}{\gamma})v_{\mathrm{s}}'} & v_{\mathrm{s}}k_{y} & v_{\mathrm{s}}k_{z}\end{array}\right).
\end{align*}
Similar to the operator $\mathcal{H}$, the symbol $H$ is also decomposed into $H_{0},H_{1}$, and $H_{2}$. $H_{0}$ is the asymptotic symbol that depends only on $v_{\mathrm{A}}$ and $v_{\mathrm{s}}$, but not their derivatives. It is the same as Eq.~(\ref{eq:symbol_H0}). $H_{1}$ depends on first order derivatives of $v_{\mathrm{A}}$ or $v_{\mathrm{s}}$, and $H_{2}$ depend on second order derivatives of $v_{\mathrm{A}}$ or $v_{\mathrm{s}}$. Clearly, the dispersion relation of $H$ is much more complicated than $H_{0}$.

Finally, for symbol $H(k_{x},k_{y},k_{z})$, the time reversal operator is complex conjugation and flipping the sign of $k_{x}$, i.e., $\mathcal{T}H(k_{x},k_{y},k_{z})\mathcal{T}=H^{*}(-k_{x},k_{y},k_{z})$. This is because we have fixed $k_{z}$ , and $k_{y}$ is treated as a control parameter. Using this property, we find that $H$ is also PT-symmetric with the same $\mathcal{P}$ used for $\mathcal{H},$ i.e., $\mathcal{P}=\mathrm{diag}(-1,1,1,-1,1,1,1)$.

Now, we approximate the Hall MHD model by a two-band non-Hermitian Hamiltonian near the degeneracy (Weyl) point between the Alfv\'{e}n wave and slow wave as determined by $H_{0}$. The index (topological charge) of the Weyl point will be calculated from the two-band approximation of $H_{0}$, and the spectral flow is explicitly solved for. According to the naming convention defined by the inequality (\ref{eq:vpslow}), the Alfv\'{e}n wave and slow wave are switched at the Weyl point. But in the neighborhood of the Weyl point, the two branches are physically the Alfv\'{e}n wave and the sound wave. Therefore, a proper name for this edge mode is topological Alfv\'{e}n-sound wave (TASW).

Assume $\zeta_{0}=v_{\mathrm{s0}}/v_{\mathrm{A0}}<1$ at the Weyl point. The location of the Weyl point in the momentum space is found to be $(k_{x},k_{y},k_{z})=\left(0,0,\left(1/\zeta_{0}-\zeta_{0}\right)/d_{i}\right)$ with eigenvalue $\omega_{0}=(v_{\mathrm{A0}}^{2}-v_{\mathrm{s0}}^{2})/(d_{\mathrm{i}}v_{\mathrm{A0}})$. At the Weyl point, the two unit eigenvectors are 
\begin{align*}
\Psi_{1}&=\dfrac{1}{\sqrt{2}}(0,0,1,0,0,0,1)^{\intercal},\\
\Psi_{2}&=\dfrac{1}{\sqrt{2\left(v_{\mathrm{A0}}^{2}+v_{\mathrm{s0}}^{2}\right)}}(-\mathrm{i}v_{\mathrm{A0}},-v_{\mathrm{A0}},0,\mathrm{i}v_{\mathrm{s0}},v_{\mathrm{s0}},0,0)^{\intercal}.
\end{align*}
Since PT symmetry is unbroken at the Weyl point, $\Psi_{1,2}$ are also the eigenvectors of the $\mathcal{PT}$ operator with eigenvalue 1, namely $\mathcal{PT}\Psi_{i}=\Psi_{i}$.

In the $(k_{x},k_{y},x)$ space near the Weyl point, we can expand the inhomogeneous equilibrium field, $v_{\mathrm{A}}(x)=v_{\mathrm{A0}}+v_{\mathrm{A0}}'x$, $v_{\mathrm{s}}(x)=v_{\mathrm{s0}}+v_{\mathrm{s0}}'x$, and approximate $H_{0}$ by the following two-band operator, 
\begin{align}
& M_{0}\doteq \boldsymbol{\Psi}^{\dagger}
\Bigg[v_{\mathrm{A}}(x)\dfrac{\partial H_{0}}{\partial v_{\mathrm{A}}}+v_{\mathrm{s}}(x)\dfrac{\partial H_{0}}{\partial v_{\mathrm{s}}} \\
& \qquad\qquad\quad + k_{y}\dfrac{\partial H_{0}}{\partial k_{y}}+k_{x}\dfrac{\partial H_{0}}{\partial k_{x}}-\omega_{0}\mathrm{I}_{2}\Bigg]
\boldsymbol{\Psi}\\[5pt]
=& \begin{pmatrix}
    \dfrac{v_{\mathrm{A0}}^{2}-v_{\mathrm{s0}}^{2}}{d_{\mathrm{i}}v_{\mathrm{A0}}}\dfrac{v_{\mathrm{s0}}'}{v_{\mathrm{s0}}}x - \omega_0 & -\dfrac{v_{\mathrm{A0}}v_{\mathrm{s0}}(k_{y}+\mathrm{i}k_{x})}{2\sqrt{v_{\mathrm{A0}}^{2}+v_{\mathrm{s0}}^{2}}}\\[15pt]
    -\dfrac{v_{\mathrm{A0}}v_{\mathrm{s0}}(k_{y}-\mathrm{i}k_{x})}{2\sqrt{v_{\mathrm{A0}}^{2}+v_{\mathrm{s0}}^{2}}} & \dfrac{v_{\mathrm{A0}}^{2}-v_{\mathrm{s0}}^{2}}{d_{\mathrm{i}}v_{\mathrm{A0}}}\dfrac{v_{\mathrm{A0}}'}{v_{\mathrm{A0}}}x  - \omega_0
\end{pmatrix}
\label{eq:two_band_symbol_0}
\end{align}
Here, $\boldsymbol{\Psi}=(\Psi_{1},\Psi_{2})^{\intercal}$, and the $\omega_{0}I_{2}$ term can be ignored. All derivatives of $H_{0}$ are evaluated at $(k_{x},k_{y},x)=(0,0,0)$. The topological charge of Weyl point is defined to be the Chern number of the slow wave bundle over a 2D surface in the $(k_{x},k_{y},x)$ space surrounding the Weyl point \cite{faure2023manifestation,qin2022topological}. Without loss of generality, assume $v_{\mathrm{A0}}'<0$, and thus $v_{\mathrm{s0}}'>0$ from the pressure balance in Eq.~(\ref{eq:equilibrium_constrain}). Let $\alpha\doteq1-v_{\mathrm{A0}}'v_{\mathrm{s0}}/(2v_{\mathrm{s0}}'v_{\mathrm{A0}})>0$. In terms of the following scaled variables, 
\begin{align}
&\tilde{x}\doteq\frac{v_{\mathrm{A0}}^{2}-v_{\mathrm{s0}}^{2}}{2d_{\mathrm{i}}v_{\mathrm{A0}}v_{\mathrm{s0}}}v_{\mathrm{s0}}'\alpha x, \\[3pt]
&\tilde{k}_{x}\doteq\frac{v_{\mathrm{A0}}v_{\mathrm{s0}}}{2\sqrt{v_{\mathrm{A0}}^{2}+v_{\mathrm{s0}}^{2}}}k_{x},
\quad\tilde{k}_{y}\doteq\frac{v_{\mathrm{A0}}v_{\mathrm{s0}}}{2\sqrt{v_{\mathrm{A0}}^{2}+v_{\mathrm{s0}}^{2}}}k_{y},
\end{align}
$M_{0}$ is simplified to 
\begin{align}
M_{0}=\begin{pmatrix}2\tilde{x}/\alpha & -\tilde{k}_{y}-\mathrm{i}\tilde{k}_{x}\\
-\tilde{k}_{y}+\mathrm{i}\tilde{k}_{x} & -2(\alpha-1)\tilde{x}/\alpha
\end{pmatrix},
\end{align}
Its eigenvalues and eigenvectors are 
\begin{align}
\omega_{\pm} & =\tilde{x}\left(\dfrac{2}{\alpha}-1\right)\pm\sqrt{\tilde{k}_{x}^{2}+\tilde{k}_{y}^{2}+\tilde{x}^{2}},\\[5pt]
\Psi_{\pm} & =\left[\dfrac{\tilde{x}\pm\sqrt{\tilde{k}_{x}^{2}+\tilde{k}_{y}^{2}+\tilde{x}^{2}}}{\mathrm{i}{\tilde{k}_{x}}-{\tilde{k}_{y}}},1\right]^{\intercal}.
\end{align}

On an infinitesimal 2D sphere in the $(\tilde{x},\tilde{k}_{x},\tilde{k}_{y})$ space centered at $(\tilde{x},\tilde{k}_{x},\tilde{k}_{y})=(0,0,0)$, the unit eigenvectors can be written in the spherical coordinates as 
\begin{align}
\Psi_{+}=\begin{pmatrix}\cos\frac{\theta}{2}\\[5pt]
\mathrm{i}\sin\frac{\theta}{2}e^{\mathrm{i}\varphi}
\end{pmatrix},\quad\Psi_{-}=\begin{pmatrix}\sin\frac{\theta}{2}\\[5pt]
-\mathrm{i}\cos\frac{\theta}{2}e^{\mathrm{i}\varphi}
\end{pmatrix}.
\end{align}
Here, the spherical coordinate is defined by $(\tilde{x},\tilde{k}_{x},\tilde{k}_{y})=(\cos\theta,\sin\theta\cos\varphi,\sin\theta\sin\varphi)$, $\theta\in[0,\pi]$, $\varphi\in[0,2\pi)$.

The Chern numbers for $\Psi_{\pm}$ are 
\begin{align*}
C_{\pm} &=\dfrac{\mathrm{1}}{2\pi}\iint\Big[\partial_{\theta}(\mathrm{i}\Psi_{\pm}^{\dagger}\partial_{\varphi}\Psi_{\pm})-\partial_{\varphi}(\mathrm{i}\Psi_{\pm}^{\dagger}\partial_{\theta}\Psi_{\pm})\Big]\mathrm{d}\varphi\,\mathrm{d}\theta \\ 
&=\mp1.
\end{align*}

Next, we derive the TASW using the two-band approximation of non-Hermitian Hall MHD model in the inhomogeneous region. To the lowest order, $M=M_{0}+M_{1}$, where $M_{0}$ is specified by Eq.~(\ref{eq:two_band_symbol_0}), and 
\begin{widetext}
    \begin{align}
        M_{1} & \doteq\boldsymbol{\Psi}^{\dagger}\left.H_{1}\right|_{x,k_{x},k_{y}=0}\boldsymbol{\Psi}=\begin{pmatrix}0 & \dfrac{v_{\mathrm{A0}}v_{\mathrm{s0}}'(1-4/\gamma)-2v_{\mathrm{A0}}'v_{\mathrm{s0}}}{4\sqrt{v_{\mathrm{A0}}^{2}+v_{\mathrm{s0}}^{2}}}\\
        \dfrac{v_{\mathrm{A0}}v_{\mathrm{s0}}'}{4\sqrt{v_{\mathrm{A0}}^{2}+v_{\mathrm{s0}}^{2}}} & 0
        \end{pmatrix}\\
         & =\dfrac{1}{4\sqrt{v_{\mathrm{A0}}^{2}+v_{\mathrm{s0}}^{2}}}\left[\Big((1-2/\gamma)v_{\mathrm{A0}}v_{\mathrm{s0}}'-v_{\mathrm{A0}}'v_{\mathrm{s0}}\Big)\sigma^{x}-\mathrm{i}\Big(2v_{\mathrm{A0}}v_{\mathrm{s0}}'/\gamma+v_{\mathrm{A0}}'v_{\mathrm{s0}}\Big)\sigma^{y}\right]
    \end{align}
\end{widetext}

The corresponding $2\times2$ operator $\mathcal{M}$ can be obtained by simply replacing $k_{x}\to-\mathrm{i}\partial_{x}$ in $M.$ Since the inhomogeneous part $M_{1}$ does not depend on $k_{x}$, we have $\mathcal{M}_{1}=M_{1}$. At this point, we are ready to verify that in $\mathcal{M}=\mathcal{M}_{0}+\mathcal{M}_{1}$ the anti-Hermitian term is proportional to $\sigma^{y}$, which is the result of PT symmetry.

Finally, we show that the non-Hermitian $\mathcal{M}$ is similar to a tilted Dirac cone, which is Hermitian and admits a spectral flow. Let 
\begin{align}
&\tilde{k}_{y}\doteq k_{y}-\dfrac{1}{2}\left[(1-2/\gamma)\dfrac{v_{\mathrm{s0}}'}{v_{\mathrm{s0}}}-\dfrac{v_{\mathrm{A0}}'}{v_{\mathrm{A0}}}\right],\\
&\delta\doteq\dfrac{1}{\gamma}\dfrac{v_{\mathrm{s0}}'}{v_{\mathrm{s0}}}-\dfrac{1}{2}\dfrac{v_{\mathrm{A0}}'}{v_{\mathrm{A0}}}.
\end{align}
Perform a similarity transformation on $\mathcal{M},$ 
\begin{align}
\tilde{\mathcal{M}} & \doteq e^{\delta x}\mathcal{M}e^{-\delta x} \\
& =\begin{pmatrix}
    \dfrac{v_{\mathrm{A0}}^{2}-v_{\mathrm{s0}}^{2}}{d_{\mathrm{i}}v_{\mathrm{A0}}}\dfrac{v_{\mathrm{s0}}'}{v_{\mathrm{s0}}}x & -\dfrac{v_{\mathrm{A0}}v_{\mathrm{s0}}(\tilde{k}_{y}+\partial_{x})}{2\sqrt{v_{\mathrm{A0}}^{2}+v_{\mathrm{s0}}^{2}}}\\[15pt]
    -\dfrac{v_{\mathrm{A0}}v_{\mathrm{s0}}(\tilde{k}_{y}-\partial_{x})}{2\sqrt{v_{\mathrm{A0}}^{2}+v_{\mathrm{s0}}^{2}}} & \dfrac{v_{\mathrm{A0}}^{2}-v_{\mathrm{s0}}^{2}}{d_{\mathrm{i}}v_{\mathrm{A0}}}\dfrac{v_{\mathrm{A0}}'}{v_{\mathrm{A0}}}x
\end{pmatrix}.
\end{align}
Note that $\tilde{\mathcal{M}}$ has the same format of the operator $\mathcal{M}_{0}=\mathscr{W}^{-1}(M_{0})$, and it can be written as 
\begin{align}
\tilde{\mathcal{M}}=c_{1}c_{2}\begin{pmatrix}\bar{x}/\kappa & -\bar{k}_{y}-\partial_{\bar{x}}\\
-\bar{k}_{y}+\partial_{\bar{x}} & -\kappa\bar{x}
\end{pmatrix},\label{eq:M-tiltedDirac}
\end{align}
where 
\begin{align}
c_{1}&\doteq\dfrac{v_{\mathrm{A0}}^{2}-v_{\mathrm{s0}}^{2}}{d_{\mathrm{i}}v_{\mathrm{A0}}}\dfrac{v_{\mathrm{s0}}'}{v_{\mathrm{s0}}}>0, \\
c_{2}&\doteq\left(\dfrac{v_{\mathrm{A0}}v_{\mathrm{s0}}\kappa}{2c_{1}\sqrt{v_{\mathrm{A0}}^{2}+v_{\mathrm{s0}}^{2}}}\right){}^{1/2},\\
\bar{x}&\doteq\dfrac{\kappa}{c_{2}}x, \quad 
\bar{k}_{y}\doteq\dfrac{c_{2}}{\kappa}k_{y},\\
\kappa&\doteq\left(-\dfrac{v_{\mathrm{A0}}'/v_{\mathrm{A0}}}{v_{\mathrm{s0}}'/v_{\mathrm{s0}}}\right)^{1/2}>0.
\end{align}
Equation (\ref{eq:M-tiltedDirac}) is the tilted Dirac cone described in Appendix~\ref{sec:analytical_spectral_flow}. It admits one spectral flow whose index is identical to the topological charge at the Weyl point \cite{qin2022topological}.

\bibliography{hmhd.bbl}

\begin{thebibliography}{72}%
\makeatletter
\providecommand \@ifxundefined [1]{%
 \@ifx{#1\undefined}
}%
\providecommand \@ifnum [1]{%
 \ifnum #1\expandafter \@firstoftwo
 \else \expandafter \@secondoftwo
 \fi
}%
\providecommand \@ifx [1]{%
 \ifx #1\expandafter \@firstoftwo
 \else \expandafter \@secondoftwo
 \fi
}%
\providecommand \natexlab [1]{#1}%
\providecommand \enquote  [1]{``#1''}%
\providecommand \bibnamefont  [1]{#1}%
\providecommand \bibfnamefont [1]{#1}%
\providecommand \citenamefont [1]{#1}%
\providecommand \href@noop [0]{\@secondoftwo}%
\providecommand \href [0]{\begingroup \@sanitize@url \@href}%
\providecommand \@href[1]{\@@startlink{#1}\@@href}%
\providecommand \@@href[1]{\endgroup#1\@@endlink}%
\providecommand \@sanitize@url [0]{\catcode `\\12\catcode `\$12\catcode
  `\&12\catcode `\#12\catcode `\^12\catcode `\_12\catcode `\%12\relax}%
\providecommand \@@startlink[1]{}%
\providecommand \@@endlink[0]{}%
\providecommand \url  [0]{\begingroup\@sanitize@url \@url }%
\providecommand \@url [1]{\endgroup\@href {#1}{\urlprefix }}%
\providecommand \urlprefix  [0]{URL }%
\providecommand \Eprint [0]{\href }%
\providecommand \doibase [0]{https://doi.org/}%
\providecommand \selectlanguage [0]{\@gobble}%
\providecommand \bibinfo  [0]{\@secondoftwo}%
\providecommand \bibfield  [0]{\@secondoftwo}%
\providecommand \translation [1]{[#1]}%
\providecommand \BibitemOpen [0]{}%
\providecommand \bibitemStop [0]{}%
\providecommand \bibitemNoStop [0]{.\EOS\space}%
\providecommand \EOS [0]{\spacefactor3000\relax}%
\providecommand \BibitemShut  [1]{\csname bibitem#1\endcsname}%
\let\auto@bib@innerbib\@empty
\bibitem [{\citenamefont {Thouless}\ \emph {et~al.}(1982)\citenamefont
  {Thouless}, \citenamefont {Kohmoto}, \citenamefont {Nightingale},\ and\
  \citenamefont {den Nijs}}]{thouless1982quantized}%
  \BibitemOpen
  \bibfield  {author} {\bibinfo {author} {\bibfnamefont {D.~J.}\ \bibnamefont
  {Thouless}}, \bibinfo {author} {\bibfnamefont {M.}~\bibnamefont {Kohmoto}},
  \bibinfo {author} {\bibfnamefont {M.~P.}\ \bibnamefont {Nightingale}},\ and\
  \bibinfo {author} {\bibfnamefont {M.}~\bibnamefont {den Nijs}},\ }\bibfield
  {title} {\bibinfo {title} {Quantized {Hall} conductance in a two-dimensional
  periodic potential},\ }\href {https://doi.org/10.1103/PhysRevLett.49.405}
  {\bibfield  {journal} {\bibinfo  {journal} {Phys. Rev. Lett.}\ }\textbf
  {\bibinfo {volume} {49}},\ \bibinfo {pages} {405} (\bibinfo {year}
  {1982})}\BibitemShut {NoStop}%
\bibitem [{\citenamefont {Simon}(1983)}]{simon1983holonomy}%
  \BibitemOpen
  \bibfield  {author} {\bibinfo {author} {\bibfnamefont {B.}~\bibnamefont
  {Simon}},\ }\bibfield  {title} {\bibinfo {title} {{Holonomy, the quantum
  adiabatic theorem, and Berry's phase}},\ }\href
  {https://doi.org/10.1103/PhysRevLett.51.2167} {\bibfield  {journal} {\bibinfo
   {journal} {Phys. Rev. Lett.}\ }\textbf {\bibinfo {volume} {51}},\ \bibinfo
  {pages} {2167} (\bibinfo {year} {1983})}\BibitemShut {NoStop}%
\bibitem [{\citenamefont {Hasan}\ and\ \citenamefont
  {Kane}(2010)}]{hasan2010colloquium}%
  \BibitemOpen
  \bibfield  {author} {\bibinfo {author} {\bibfnamefont {M.~Z.}\ \bibnamefont
  {Hasan}}\ and\ \bibinfo {author} {\bibfnamefont {C.~L.}\ \bibnamefont
  {Kane}},\ }\bibfield  {title} {\bibinfo {title} {Colloquium: topological
  insulators},\ }\href {https://doi.org/10.1103/RevModPhys.82.3045} {\bibfield
  {journal} {\bibinfo  {journal} {Rev. Mod. Phys.}\ }\textbf {\bibinfo {volume}
  {82}},\ \bibinfo {pages} {3045} (\bibinfo {year} {2010})}\BibitemShut
  {NoStop}%
\bibitem [{\citenamefont {Qi}\ and\ \citenamefont
  {Zhang}(2011)}]{qi2011topological}%
  \BibitemOpen
  \bibfield  {author} {\bibinfo {author} {\bibfnamefont {X.-L.}\ \bibnamefont
  {Qi}}\ and\ \bibinfo {author} {\bibfnamefont {S.-C.}\ \bibnamefont {Zhang}},\
  }\bibfield  {title} {\bibinfo {title} {Topological insulators and
  superconductors},\ }\href {https://doi.org/10.1103/RevModPhys.83.1057}
  {\bibfield  {journal} {\bibinfo  {journal} {Rev. Mod. Phys.}\ }\textbf
  {\bibinfo {volume} {83}},\ \bibinfo {pages} {1057} (\bibinfo {year}
  {2011})}\BibitemShut {NoStop}%
\bibitem [{\citenamefont {Armitage}\ \emph {et~al.}(2018)\citenamefont
  {Armitage}, \citenamefont {Mele},\ and\ \citenamefont
  {Vishwanath}}]{armitage2018weyl}%
  \BibitemOpen
  \bibfield  {author} {\bibinfo {author} {\bibfnamefont {N.~P.}\ \bibnamefont
  {Armitage}}, \bibinfo {author} {\bibfnamefont {E.~J.}\ \bibnamefont {Mele}},\
  and\ \bibinfo {author} {\bibfnamefont {A.}~\bibnamefont {Vishwanath}},\
  }\bibfield  {title} {\bibinfo {title} {{Weyl and Dirac semimetals in
  three-dimensional solids}},\ }\href
  {https://doi.org/10.1103/RevModPhys.90.015001} {\bibfield  {journal}
  {\bibinfo  {journal} {Rev. Mod. Phys.}\ }\textbf {\bibinfo {volume} {90}},\
  \bibinfo {pages} {015001} (\bibinfo {year} {2018})}\BibitemShut {NoStop}%
\bibitem [{\citenamefont {Kitaev}(2009)}]{kitaev2009periodic}%
  \BibitemOpen
  \bibfield  {author} {\bibinfo {author} {\bibfnamefont {A.}~\bibnamefont
  {Kitaev}},\ }\bibfield  {title} {\bibinfo {title} {Periodic table for
  topological insulators and superconductors},\ }\href
  {https://doi.org/10.1063/1.3149495} {\bibfield  {journal} {\bibinfo
  {journal} {AIP Conf. Proc.}\ }\textbf {\bibinfo {volume} {1134}},\ \bibinfo
  {pages} {22} (\bibinfo {year} {2009})}\BibitemShut {NoStop}%
\bibitem [{\citenamefont {Chiu}\ \emph {et~al.}(2016)\citenamefont {Chiu},
  \citenamefont {Teo}, \citenamefont {Schnyder},\ and\ \citenamefont
  {Ryu}}]{chiu2016classification}%
  \BibitemOpen
  \bibfield  {author} {\bibinfo {author} {\bibfnamefont {C.-K.}\ \bibnamefont
  {Chiu}}, \bibinfo {author} {\bibfnamefont {J.~C.~Y.}\ \bibnamefont {Teo}},
  \bibinfo {author} {\bibfnamefont {A.~P.}\ \bibnamefont {Schnyder}},\ and\
  \bibinfo {author} {\bibfnamefont {S.}~\bibnamefont {Ryu}},\ }\bibfield
  {title} {\bibinfo {title} {Classification of topological quantum matter with
  symmetries},\ }\href {https://doi.org/10.1103/RevModPhys.88.035005}
  {\bibfield  {journal} {\bibinfo  {journal} {Rev. Mod. Phys.}\ }\textbf
  {\bibinfo {volume} {88}},\ \bibinfo {pages} {035005} (\bibinfo {year}
  {2016})}\BibitemShut {NoStop}%
\bibitem [{\citenamefont {Haldane}\ and\ \citenamefont
  {Raghu}(2008)}]{haldane2008possible}%
  \BibitemOpen
  \bibfield  {author} {\bibinfo {author} {\bibfnamefont {F.~D.~M.}\
  \bibnamefont {Haldane}}\ and\ \bibinfo {author} {\bibfnamefont
  {S.}~\bibnamefont {Raghu}},\ }\bibfield  {title} {\bibinfo {title} {Possible
  realization of directional optical waveguides in photonic crystals with
  broken time-reversal symmetry},\ }\href
  {https://doi.org/10.1103/PhysRevLett.100.013904} {\bibfield  {journal}
  {\bibinfo  {journal} {Phys. Rev. Lett.}\ }\textbf {\bibinfo {volume} {100}},\
  \bibinfo {pages} {013904} (\bibinfo {year} {2008})}\BibitemShut {NoStop}%
\bibitem [{\citenamefont {Raghu}\ and\ \citenamefont
  {Haldane}(2008)}]{raghu2008analogs}%
  \BibitemOpen
  \bibfield  {author} {\bibinfo {author} {\bibfnamefont {S.}~\bibnamefont
  {Raghu}}\ and\ \bibinfo {author} {\bibfnamefont {F.~D.~M.}\ \bibnamefont
  {Haldane}},\ }\bibfield  {title} {\bibinfo {title} {{Analogs of
  quantum-Hall-effect edge states in photonic crystals}},\ }\href
  {https://doi.org/10.1103/PhysRevA.78.033834} {\bibfield  {journal} {\bibinfo
  {journal} {Phys. Rev. A}\ }\textbf {\bibinfo {volume} {78}},\ \bibinfo
  {pages} {033834} (\bibinfo {year} {2008})}\BibitemShut {NoStop}%
\bibitem [{\citenamefont {Ozawa}\ \emph {et~al.}(2019)\citenamefont {Ozawa},
  \citenamefont {Price}, \citenamefont {Amo}, \citenamefont {Goldman},
  \citenamefont {Hafezi}, \citenamefont {Lu}, \citenamefont {Rechtsman},
  \citenamefont {Schuster}, \citenamefont {Simon}, \citenamefont {Zilberberg}
  \emph {et~al.}}]{ozawa2019topological}%
  \BibitemOpen
  \bibfield  {author} {\bibinfo {author} {\bibfnamefont {T.}~\bibnamefont
  {Ozawa}}, \bibinfo {author} {\bibfnamefont {H.~M.}\ \bibnamefont {Price}},
  \bibinfo {author} {\bibfnamefont {A.}~\bibnamefont {Amo}}, \bibinfo {author}
  {\bibfnamefont {N.}~\bibnamefont {Goldman}}, \bibinfo {author} {\bibfnamefont
  {M.}~\bibnamefont {Hafezi}}, \bibinfo {author} {\bibfnamefont
  {L.}~\bibnamefont {Lu}}, \bibinfo {author} {\bibfnamefont {M.~C.}\
  \bibnamefont {Rechtsman}}, \bibinfo {author} {\bibfnamefont {D.}~\bibnamefont
  {Schuster}}, \bibinfo {author} {\bibfnamefont {J.}~\bibnamefont {Simon}},
  \bibinfo {author} {\bibfnamefont {O.}~\bibnamefont {Zilberberg}}, \emph
  {et~al.},\ }\bibfield  {title} {\bibinfo {title} {Topological photonics},\
  }\href {https://doi.org/10.1103/RevModPhys.91.015006} {\bibfield  {journal}
  {\bibinfo  {journal} {Rev. Mod. Phys.}\ }\textbf {\bibinfo {volume} {91}},\
  \bibinfo {pages} {015006} (\bibinfo {year} {2019})}\BibitemShut {NoStop}%
\bibitem [{\citenamefont {Prodan}\ and\ \citenamefont
  {Prodan}(2009)}]{prodan2009topological}%
  \BibitemOpen
  \bibfield  {author} {\bibinfo {author} {\bibfnamefont {E.}~\bibnamefont
  {Prodan}}\ and\ \bibinfo {author} {\bibfnamefont {C.}~\bibnamefont
  {Prodan}},\ }\bibfield  {title} {\bibinfo {title} {Topological phonon modes
  and their role in dynamic instability of microtubules},\ }\href
  {https://doi.org/10.1103/PhysRevLett.103.248101} {\bibfield  {journal}
  {\bibinfo  {journal} {Phys. Rev. Lett.}\ }\textbf {\bibinfo {volume} {103}},\
  \bibinfo {pages} {248101} (\bibinfo {year} {2009})}\BibitemShut {NoStop}%
\bibitem [{\citenamefont {Zhang}\ \emph {et~al.}(2010)\citenamefont {Zhang},
  \citenamefont {Ren}, \citenamefont {Wang},\ and\ \citenamefont
  {Li}}]{zhang2010topological}%
  \BibitemOpen
  \bibfield  {author} {\bibinfo {author} {\bibfnamefont {L.}~\bibnamefont
  {Zhang}}, \bibinfo {author} {\bibfnamefont {J.}~\bibnamefont {Ren}}, \bibinfo
  {author} {\bibfnamefont {J.-S.}\ \bibnamefont {Wang}},\ and\ \bibinfo
  {author} {\bibfnamefont {B.}~\bibnamefont {Li}},\ }\bibfield  {title}
  {\bibinfo {title} {{Topological nature of the phonon Hall effect}},\ }\href
  {https://doi.org/10.1103/PhysRevLett.105.225901} {\bibfield  {journal}
  {\bibinfo  {journal} {Phys. Rev. Lett.}\ }\textbf {\bibinfo {volume} {105}},\
  \bibinfo {pages} {225901} (\bibinfo {year} {2010})}\BibitemShut {NoStop}%
\bibitem [{\citenamefont {Yang}\ \emph {et~al.}(2015)\citenamefont {Yang},
  \citenamefont {Gao}, \citenamefont {Shi}, \citenamefont {Lin}, \citenamefont
  {Gao}, \citenamefont {Chong},\ and\ \citenamefont
  {Zhang}}]{yang2015topological}%
  \BibitemOpen
  \bibfield  {author} {\bibinfo {author} {\bibfnamefont {Z.}~\bibnamefont
  {Yang}}, \bibinfo {author} {\bibfnamefont {F.}~\bibnamefont {Gao}}, \bibinfo
  {author} {\bibfnamefont {X.}~\bibnamefont {Shi}}, \bibinfo {author}
  {\bibfnamefont {X.}~\bibnamefont {Lin}}, \bibinfo {author} {\bibfnamefont
  {Z.}~\bibnamefont {Gao}}, \bibinfo {author} {\bibfnamefont {Y.}~\bibnamefont
  {Chong}},\ and\ \bibinfo {author} {\bibfnamefont {B.}~\bibnamefont {Zhang}},\
  }\bibfield  {title} {\bibinfo {title} {Topological acoustics},\ }\href
  {https://doi.org/10.1103/PhysRevLett.114.114301} {\bibfield  {journal}
  {\bibinfo  {journal} {Phys. Rev. Lett.}\ }\textbf {\bibinfo {volume} {114}},\
  \bibinfo {pages} {114301} (\bibinfo {year} {2015})}\BibitemShut {NoStop}%
\bibitem [{\citenamefont {Wang}\ \emph {et~al.}(2015)\citenamefont {Wang},
  \citenamefont {Lu},\ and\ \citenamefont {Bertoldi}}]{wang2015topological}%
  \BibitemOpen
  \bibfield  {author} {\bibinfo {author} {\bibfnamefont {P.}~\bibnamefont
  {Wang}}, \bibinfo {author} {\bibfnamefont {L.}~\bibnamefont {Lu}},\ and\
  \bibinfo {author} {\bibfnamefont {K.}~\bibnamefont {Bertoldi}},\ }\bibfield
  {title} {\bibinfo {title} {Topological phononic crystals with one-way elastic
  edge waves},\ }\href {https://doi.org/10.1103/PhysRevLett.115.104302}
  {\bibfield  {journal} {\bibinfo  {journal} {Phys. Rev. Lett.}\ }\textbf
  {\bibinfo {volume} {115}},\ \bibinfo {pages} {104302} (\bibinfo {year}
  {2015})}\BibitemShut {NoStop}%
\bibitem [{\citenamefont {Souslov}\ \emph {et~al.}(2019)\citenamefont
  {Souslov}, \citenamefont {Dasbiswas}, \citenamefont {Fruchart}, \citenamefont
  {Vaikuntanathan},\ and\ \citenamefont {Vitelli}}]{souslov2019topological}%
  \BibitemOpen
  \bibfield  {author} {\bibinfo {author} {\bibfnamefont {A.}~\bibnamefont
  {Souslov}}, \bibinfo {author} {\bibfnamefont {K.}~\bibnamefont {Dasbiswas}},
  \bibinfo {author} {\bibfnamefont {M.}~\bibnamefont {Fruchart}}, \bibinfo
  {author} {\bibfnamefont {S.}~\bibnamefont {Vaikuntanathan}},\ and\ \bibinfo
  {author} {\bibfnamefont {V.}~\bibnamefont {Vitelli}},\ }\bibfield  {title}
  {\bibinfo {title} {Topological waves in fluids with odd viscosity},\ }\href
  {https://doi.org/10.1103/PhysRevLett.122.128001} {\bibfield  {journal}
  {\bibinfo  {journal} {Phys. Rev. Lett.}\ }\textbf {\bibinfo {volume} {122}},\
  \bibinfo {pages} {128001} (\bibinfo {year} {2019})}\BibitemShut {NoStop}%
\bibitem [{\citenamefont {Delplace}\ \emph {et~al.}(2017)\citenamefont
  {Delplace}, \citenamefont {Marston},\ and\ \citenamefont
  {Venaille}}]{delplace2017topological}%
  \BibitemOpen
  \bibfield  {author} {\bibinfo {author} {\bibfnamefont {P.}~\bibnamefont
  {Delplace}}, \bibinfo {author} {\bibfnamefont {J.~B.}\ \bibnamefont
  {Marston}},\ and\ \bibinfo {author} {\bibfnamefont {A.}~\bibnamefont
  {Venaille}},\ }\bibfield  {title} {\bibinfo {title} {Topological origin of
  equatorial waves},\ }\href {https://doi.org/10.1126/science.aan8819}
  {\bibfield  {journal} {\bibinfo  {journal} {Science}\ }\textbf {\bibinfo
  {volume} {358}},\ \bibinfo {pages} {1075} (\bibinfo {year}
  {2017})}\BibitemShut {NoStop}%
\bibitem [{\citenamefont {Perrot}\ \emph {et~al.}(2019)\citenamefont {Perrot},
  \citenamefont {Delplace},\ and\ \citenamefont
  {Venaille}}]{perrot2019topological}%
  \BibitemOpen
  \bibfield  {author} {\bibinfo {author} {\bibfnamefont {M.}~\bibnamefont
  {Perrot}}, \bibinfo {author} {\bibfnamefont {P.}~\bibnamefont {Delplace}},\
  and\ \bibinfo {author} {\bibfnamefont {A.}~\bibnamefont {Venaille}},\
  }\bibfield  {title} {\bibinfo {title} {Topological transition in stratified
  fluids},\ }\href {https://doi.org/10.1038/s41567-019-0561-1} {\bibfield
  {journal} {\bibinfo  {journal} {Nat. Phys.}\ }\textbf {\bibinfo {volume}
  {15}},\ \bibinfo {pages} {781} (\bibinfo {year} {2019})}\BibitemShut
  {NoStop}%
\bibitem [{\citenamefont {Perez}\ \emph {et~al.}(2022)\citenamefont {Perez},
  \citenamefont {Delplace},\ and\ \citenamefont
  {Venaille}}]{perez2022unidirectional}%
  \BibitemOpen
  \bibfield  {author} {\bibinfo {author} {\bibfnamefont {N.}~\bibnamefont
  {Perez}}, \bibinfo {author} {\bibfnamefont {P.}~\bibnamefont {Delplace}},\
  and\ \bibinfo {author} {\bibfnamefont {A.}~\bibnamefont {Venaille}},\
  }\bibfield  {title} {\bibinfo {title} {{Unidirectional modes induced by
  nontraditional Coriolis force in stratified fluids}},\ }\href
  {https://doi.org/10.1103/PhysRevLett.128.184501} {\bibfield  {journal}
  {\bibinfo  {journal} {Phys. Rev. Lett.}\ }\textbf {\bibinfo {volume} {128}},\
  \bibinfo {pages} {184501} (\bibinfo {year} {2022})}\BibitemShut {NoStop}%
\bibitem [{\citenamefont {Zhu}\ \emph {et~al.}(2023)\citenamefont {Zhu},
  \citenamefont {Li},\ and\ \citenamefont {Marston}}]{Zhu2023}%
  \BibitemOpen
  \bibfield  {author} {\bibinfo {author} {\bibfnamefont {Z.}~\bibnamefont
  {Zhu}}, \bibinfo {author} {\bibfnamefont {C.}~\bibnamefont {Li}},\ and\
  \bibinfo {author} {\bibfnamefont {J.~B.}\ \bibnamefont {Marston}},\
  }\bibfield  {title} {\bibinfo {title} {Topology of rotating stratified fluids
  with and without background shear flow},\ }\href
  {https://doi.org/10.1103/PhysRevResearch.5.033191} {\bibfield  {journal}
  {\bibinfo  {journal} {Phys. Rev. Res.}\ }\textbf {\bibinfo {volume} {5}},\
  \bibinfo {pages} {033191} (\bibinfo {year} {2023})}\BibitemShut {NoStop}%
\bibitem [{\citenamefont {Tong}(2023)}]{tong2023gauge}%
  \BibitemOpen
  \bibfield  {author} {\bibinfo {author} {\bibfnamefont {D.}~\bibnamefont
  {Tong}},\ }\bibfield  {title} {\bibinfo {title} {{A gauge theory for shallow
  water}},\ }\href {https://doi.org/10.21468/SciPostPhys.14.5.102} {\bibfield
  {journal} {\bibinfo  {journal} {SciPost Phys.}\ }\textbf {\bibinfo {volume}
  {14}},\ \bibinfo {pages} {102} (\bibinfo {year} {2023})}\BibitemShut
  {NoStop}%
\bibitem [{\citenamefont {Gao}\ \emph {et~al.}(2016)\citenamefont {Gao},
  \citenamefont {Yang}, \citenamefont {Lawrence}, \citenamefont {Fang},
  \citenamefont {B{\'e}ri},\ and\ \citenamefont {Zhang}}]{gao2016photonic}%
  \BibitemOpen
  \bibfield  {author} {\bibinfo {author} {\bibfnamefont {W.}~\bibnamefont
  {Gao}}, \bibinfo {author} {\bibfnamefont {B.}~\bibnamefont {Yang}}, \bibinfo
  {author} {\bibfnamefont {M.}~\bibnamefont {Lawrence}}, \bibinfo {author}
  {\bibfnamefont {F.}~\bibnamefont {Fang}}, \bibinfo {author} {\bibfnamefont
  {B.}~\bibnamefont {B{\'e}ri}},\ and\ \bibinfo {author} {\bibfnamefont
  {S.}~\bibnamefont {Zhang}},\ }\bibfield  {title} {\bibinfo {title} {{Photonic
  Weyl degeneracies in magnetized plasma}},\ }\href
  {https://doi.org/10.1038/ncomms12435} {\bibfield  {journal} {\bibinfo
  {journal} {Nat. Commun.}\ }\textbf {\bibinfo {volume} {7}},\ \bibinfo {pages}
  {12435} (\bibinfo {year} {2016})}\BibitemShut {NoStop}%
\bibitem [{\citenamefont {Parker}\ \emph {et~al.}(2020)\citenamefont {Parker},
  \citenamefont {Marston}, \citenamefont {Tobias},\ and\ \citenamefont
  {Zhu}}]{parker2020topological}%
  \BibitemOpen
  \bibfield  {author} {\bibinfo {author} {\bibfnamefont {J.~B.}\ \bibnamefont
  {Parker}}, \bibinfo {author} {\bibfnamefont {J.~B.}\ \bibnamefont {Marston}},
  \bibinfo {author} {\bibfnamefont {S.~M.}\ \bibnamefont {Tobias}},\ and\
  \bibinfo {author} {\bibfnamefont {Z.}~\bibnamefont {Zhu}},\ }\bibfield
  {title} {\bibinfo {title} {Topological gaseous plasmon polariton in realistic
  plasma},\ }\href {https://doi.org/10.1103/PhysRevLett.124.195001} {\bibfield
  {journal} {\bibinfo  {journal} {Phys. Rev. Lett.}\ }\textbf {\bibinfo
  {volume} {124}},\ \bibinfo {pages} {195001} (\bibinfo {year}
  {2020})}\BibitemShut {NoStop}%
\bibitem [{\citenamefont {Fu}\ and\ \citenamefont
  {Qin}(2021)}]{fu2021topological}%
  \BibitemOpen
  \bibfield  {author} {\bibinfo {author} {\bibfnamefont {Y.}~\bibnamefont
  {Fu}}\ and\ \bibinfo {author} {\bibfnamefont {H.}~\bibnamefont {Qin}},\
  }\bibfield  {title} {\bibinfo {title} {Topological phases and bulk-edge
  correspondence of magnetized cold plasmas},\ }\href
  {https://doi.org/10.1038/s41467-021-24189-3} {\bibfield  {journal} {\bibinfo
  {journal} {Nat. Commun.}\ }\textbf {\bibinfo {volume} {12}},\ \bibinfo
  {pages} {3924} (\bibinfo {year} {2021})}\BibitemShut {NoStop}%
\bibitem [{\citenamefont {Fu}\ and\ \citenamefont
  {Qin}(2022)}]{fu2022dispersion}%
  \BibitemOpen
  \bibfield  {author} {\bibinfo {author} {\bibfnamefont {Y.}~\bibnamefont
  {Fu}}\ and\ \bibinfo {author} {\bibfnamefont {H.}~\bibnamefont {Qin}},\
  }\bibfield  {title} {\bibinfo {title} {{The dispersion and propagation of
  topological Langmuir-cyclotron waves in cold magnetized plasmas}},\ }\href
  {https://doi.org/10.1017/S0022377822000629} {\bibfield  {journal} {\bibinfo
  {journal} {J. Plasma Phys.}\ }\textbf {\bibinfo {volume} {88}},\ \bibinfo
  {pages} {835880401} (\bibinfo {year} {2022})}\BibitemShut {NoStop}%
\bibitem [{\citenamefont {{Palmerduca}}\ and\ \citenamefont
  {{Qin}}(2023)}]{palmerduca2023photon}%
  \BibitemOpen
  \bibfield  {author} {\bibinfo {author} {\bibfnamefont {E.}~\bibnamefont
  {{Palmerduca}}}\ and\ \bibinfo {author} {\bibfnamefont {H.}~\bibnamefont
  {{Qin}}},\ }\bibfield  {title} {\bibinfo {title} {{Photon topology}},\ }\href
  {https://doi.org/10.48550/arXiv.2308.11147} {\bibfield  {journal} {\bibinfo
  {journal} {arXiv e-prints}\ ,\ \bibinfo {pages} {arXiv:2308.11147}} (\bibinfo
  {year} {2023})}\BibitemShut {NoStop}%
\bibitem [{\citenamefont {Bernevig}\ \emph {et~al.}(2006)\citenamefont
  {Bernevig}, \citenamefont {Hughes},\ and\ \citenamefont
  {Zhang}}]{bernevig2006quantum}%
  \BibitemOpen
  \bibfield  {author} {\bibinfo {author} {\bibfnamefont {B.~A.}\ \bibnamefont
  {Bernevig}}, \bibinfo {author} {\bibfnamefont {T.~L.}\ \bibnamefont
  {Hughes}},\ and\ \bibinfo {author} {\bibfnamefont {S.-C.}\ \bibnamefont
  {Zhang}},\ }\bibfield  {title} {\bibinfo {title} {{Quantum spin Hall effect
  and topological phase transition in HgTe quantum wells}},\ }\href
  {https://doi.org/10.1126/science.1133734} {\bibfield  {journal} {\bibinfo
  {journal} {Science}\ }\textbf {\bibinfo {volume} {314}},\ \bibinfo {pages}
  {1757} (\bibinfo {year} {2006})}\BibitemShut {NoStop}%
\bibitem [{\citenamefont {Zyuzin}\ and\ \citenamefont
  {Burkov}(2012)}]{zyuzin2012topological}%
  \BibitemOpen
  \bibfield  {author} {\bibinfo {author} {\bibfnamefont {A.~A.}\ \bibnamefont
  {Zyuzin}}\ and\ \bibinfo {author} {\bibfnamefont {A.~A.}\ \bibnamefont
  {Burkov}},\ }\bibfield  {title} {\bibinfo {title} {{Topological response in
  Weyl semimetals and the chiral anomaly}},\ }\href
  {https://doi.org/10.1103/PhysRevB.86.115133} {\bibfield  {journal} {\bibinfo
  {journal} {Phys. Rev. B}\ }\textbf {\bibinfo {volume} {86}},\ \bibinfo
  {pages} {115133} (\bibinfo {year} {2012})}\BibitemShut {NoStop}%
\bibitem [{\citenamefont {Leykam}\ \emph {et~al.}(2017)\citenamefont {Leykam},
  \citenamefont {Bliokh}, \citenamefont {Huang}, \citenamefont {Chong},\ and\
  \citenamefont {Nori}}]{leykam2017edge}%
  \BibitemOpen
  \bibfield  {author} {\bibinfo {author} {\bibfnamefont {D.}~\bibnamefont
  {Leykam}}, \bibinfo {author} {\bibfnamefont {K.~Y.}\ \bibnamefont {Bliokh}},
  \bibinfo {author} {\bibfnamefont {C.}~\bibnamefont {Huang}}, \bibinfo
  {author} {\bibfnamefont {Y.~D.}\ \bibnamefont {Chong}},\ and\ \bibinfo
  {author} {\bibfnamefont {F.}~\bibnamefont {Nori}},\ }\bibfield  {title}
  {\bibinfo {title} {{Edge modes, degeneracies, and topological numbers in
  non-Hermitian systems}},\ }\href
  {https://doi.org/10.1103/PhysRevLett.118.040401} {\bibfield  {journal}
  {\bibinfo  {journal} {Phys. Rev. Lett.}\ }\textbf {\bibinfo {volume} {118}},\
  \bibinfo {pages} {040401} (\bibinfo {year} {2017})}\BibitemShut {NoStop}%
\bibitem [{\citenamefont {Shen}\ \emph {et~al.}(2018)\citenamefont {Shen},
  \citenamefont {Zhen},\ and\ \citenamefont {Fu}}]{shen2018topological}%
  \BibitemOpen
  \bibfield  {author} {\bibinfo {author} {\bibfnamefont {H.}~\bibnamefont
  {Shen}}, \bibinfo {author} {\bibfnamefont {B.}~\bibnamefont {Zhen}},\ and\
  \bibinfo {author} {\bibfnamefont {L.}~\bibnamefont {Fu}},\ }\bibfield
  {title} {\bibinfo {title} {{Topological band theory for non-Hermitian
  Hamiltonians}},\ }\href {https://doi.org/10.1103/PhysRevLett.120.146402}
  {\bibfield  {journal} {\bibinfo  {journal} {Phys. Rev. Lett.}\ }\textbf
  {\bibinfo {volume} {120}},\ \bibinfo {pages} {146402} (\bibinfo {year}
  {2018})}\BibitemShut {NoStop}%
\bibitem [{\citenamefont {Delplace}(2022)}]{delplace2022berry}%
  \BibitemOpen
  \bibfield  {author} {\bibinfo {author} {\bibfnamefont {P.}~\bibnamefont
  {Delplace}},\ }\bibfield  {title} {\bibinfo {title} {{Berry-Chern monopoles
  and spectral flows}},\ }\href
  {https://doi.org/10.21468/SciPostPhysLectNotes.39} {\bibfield  {journal}
  {\bibinfo  {journal} {SciPost Phys. Lect. Notes}\ ,\ \bibinfo {pages} {039}}
  (\bibinfo {year} {2022})}\BibitemShut {NoStop}%
\bibitem [{\citenamefont {Qin}\ and\ \citenamefont
  {Fu}(2023)}]{qin2022topological}%
  \BibitemOpen
  \bibfield  {author} {\bibinfo {author} {\bibfnamefont {H.}~\bibnamefont
  {Qin}}\ and\ \bibinfo {author} {\bibfnamefont {Y.}~\bibnamefont {Fu}},\
  }\bibfield  {title} {\bibinfo {title} {{Topological Langmuir-cyclotron
  wave}},\ }\href {https://doi.org/10.1126/sciadv.add8041} {\bibfield
  {journal} {\bibinfo  {journal} {Sci. Adv.}\ }\textbf {\bibinfo {volume}
  {9}},\ \bibinfo {pages} {eadd8041} (\bibinfo {year} {2023})}\BibitemShut
  {NoStop}%
\bibitem [{\citenamefont {Faure}(2023)}]{faure2023manifestation}%
  \BibitemOpen
  \bibfield  {author} {\bibinfo {author} {\bibfnamefont {F.}~\bibnamefont
  {Faure}},\ }\bibfield  {title} {\bibinfo {title} {Manifestation of the
  topological index formula in quantum waves and geophysical waves},\ }\href
  {https://doi.org/10.5802/ahl.169} {\bibfield  {journal} {\bibinfo  {journal}
  {Annales Henri Lebesgue}\ }\textbf {\bibinfo {volume} {6}},\ \bibinfo {pages}
  {449} (\bibinfo {year} {2023})}\BibitemShut {NoStop}%
\bibitem [{\citenamefont {Marciani}\ and\ \citenamefont
  {Delplace}(2020)}]{marciani2020chiral}%
  \BibitemOpen
  \bibfield  {author} {\bibinfo {author} {\bibfnamefont {M.}~\bibnamefont
  {Marciani}}\ and\ \bibinfo {author} {\bibfnamefont {P.}~\bibnamefont
  {Delplace}},\ }\bibfield  {title} {\bibinfo {title} {{Chiral Maxwell waves in
  continuous media from Berry monopoles}},\ }\href
  {https://doi.org/10.1103/PhysRevA.101.023827} {\bibfield  {journal} {\bibinfo
   {journal} {Phys. Rev. A}\ }\textbf {\bibinfo {volume} {101}},\ \bibinfo
  {pages} {023827} (\bibinfo {year} {2020})}\BibitemShut {NoStop}%
\bibitem [{\citenamefont {Bergholtz}\ \emph {et~al.}(2021)\citenamefont
  {Bergholtz}, \citenamefont {Budich},\ and\ \citenamefont
  {Kunst}}]{bergholtz2021exceptional}%
  \BibitemOpen
  \bibfield  {author} {\bibinfo {author} {\bibfnamefont {E.~J.}\ \bibnamefont
  {Bergholtz}}, \bibinfo {author} {\bibfnamefont {J.~C.}\ \bibnamefont
  {Budich}},\ and\ \bibinfo {author} {\bibfnamefont {F.~K.}\ \bibnamefont
  {Kunst}},\ }\bibfield  {title} {\bibinfo {title} {{Exceptional topology of
  non-Hermitian systems}},\ }\href
  {https://doi.org/10.1103/RevModPhys.93.015005} {\bibfield  {journal}
  {\bibinfo  {journal} {Rev. Mod. Phys.}\ }\textbf {\bibinfo {volume} {93}},\
  \bibinfo {pages} {015005} (\bibinfo {year} {2021})}\BibitemShut {NoStop}%
\bibitem [{\citenamefont {Wang}\ \emph {et~al.}(2021)\citenamefont {Wang},
  \citenamefont {Zhang}, \citenamefont {Hua}, \citenamefont {Lei},
  \citenamefont {Lu},\ and\ \citenamefont {Chen}}]{wang2021topological}%
  \BibitemOpen
  \bibfield  {author} {\bibinfo {author} {\bibfnamefont {H.}~\bibnamefont
  {Wang}}, \bibinfo {author} {\bibfnamefont {X.}~\bibnamefont {Zhang}},
  \bibinfo {author} {\bibfnamefont {J.}~\bibnamefont {Hua}}, \bibinfo {author}
  {\bibfnamefont {D.}~\bibnamefont {Lei}}, \bibinfo {author} {\bibfnamefont
  {M.}~\bibnamefont {Lu}},\ and\ \bibinfo {author} {\bibfnamefont
  {Y.}~\bibnamefont {Chen}},\ }\bibfield  {title} {\bibinfo {title}
  {Topological physics of non-hermitian optics and photonics: a review},\
  }\href {https://doi.org/10.1088/2040-8986/ac2e15} {\bibfield  {journal}
  {\bibinfo  {journal} {J. Opt.}\ }\textbf {\bibinfo {volume} {23}},\ \bibinfo
  {pages} {123001} (\bibinfo {year} {2021})}\BibitemShut {NoStop}%
\bibitem [{\citenamefont {Okuma}\ and\ \citenamefont
  {Sato}(2023)}]{okuma2023non}%
  \BibitemOpen
  \bibfield  {author} {\bibinfo {author} {\bibfnamefont {N.}~\bibnamefont
  {Okuma}}\ and\ \bibinfo {author} {\bibfnamefont {M.}~\bibnamefont {Sato}},\
  }\bibfield  {title} {\bibinfo {title} {{Non-hermitian topological phenomena:
  A review}},\ }\href
  {https://doi.org/10.1146/annurev-conmatphys-040521-033133} {\bibfield
  {journal} {\bibinfo  {journal} {Annu. Rev. Condens. Matter Phys.}\ }\textbf
  {\bibinfo {volume} {14}},\ \bibinfo {pages} {83} (\bibinfo {year}
  {2023})}\BibitemShut {NoStop}%
\bibitem [{\citenamefont {Lee}(2016)}]{lee2016anomalous}%
  \BibitemOpen
  \bibfield  {author} {\bibinfo {author} {\bibfnamefont {T.~E.}\ \bibnamefont
  {Lee}},\ }\bibfield  {title} {\bibinfo {title} {{Anomalous edge state in a
  non-Hermitian lattice}},\ }\href
  {https://doi.org/10.1103/PhysRevLett.116.133903} {\bibfield  {journal}
  {\bibinfo  {journal} {Phys. Rev. Lett.}\ }\textbf {\bibinfo {volume} {116}},\
  \bibinfo {pages} {133903} (\bibinfo {year} {2016})}\BibitemShut {NoStop}%
\bibitem [{\citenamefont {Xiong}(2018)}]{xiong2018does}%
  \BibitemOpen
  \bibfield  {author} {\bibinfo {author} {\bibfnamefont {Y.}~\bibnamefont
  {Xiong}},\ }\bibfield  {title} {\bibinfo {title} {Why does bulk boundary
  correspondence fail in some non-hermitian topological models},\ }\href
  {https://doi.org/10.1088/2399-6528/aab64a} {\bibfield  {journal} {\bibinfo
  {journal} {J. Phys. Commun.}\ }\textbf {\bibinfo {volume} {2}},\ \bibinfo
  {pages} {035043} (\bibinfo {year} {2018})}\BibitemShut {NoStop}%
\bibitem [{\citenamefont {Yao}\ and\ \citenamefont {Wang}(2018)}]{yao2018edge}%
  \BibitemOpen
  \bibfield  {author} {\bibinfo {author} {\bibfnamefont {S.}~\bibnamefont
  {Yao}}\ and\ \bibinfo {author} {\bibfnamefont {Z.}~\bibnamefont {Wang}},\
  }\bibfield  {title} {\bibinfo {title} {{Edge states and topological
  invariants of non-Hermitian systems}},\ }\href
  {https://doi.org/10.1103/PhysRevLett.121.086803} {\bibfield  {journal}
  {\bibinfo  {journal} {Phys. Rev. Lett.}\ }\textbf {\bibinfo {volume} {121}},\
  \bibinfo {pages} {086803} (\bibinfo {year} {2018})}\BibitemShut {NoStop}%
\bibitem [{\citenamefont {Yao}\ \emph {et~al.}(2018)\citenamefont {Yao},
  \citenamefont {Song},\ and\ \citenamefont {Wang}}]{yao2018non}%
  \BibitemOpen
  \bibfield  {author} {\bibinfo {author} {\bibfnamefont {S.}~\bibnamefont
  {Yao}}, \bibinfo {author} {\bibfnamefont {F.}~\bibnamefont {Song}},\ and\
  \bibinfo {author} {\bibfnamefont {Z.}~\bibnamefont {Wang}},\ }\bibfield
  {title} {\bibinfo {title} {{Non-Hermitian Chern bands}},\ }\href
  {https://doi.org/10.1103/PhysRevLett.121.136802} {\bibfield  {journal}
  {\bibinfo  {journal} {Phys. Rev. Lett.}\ }\textbf {\bibinfo {volume} {121}},\
  \bibinfo {pages} {136802} (\bibinfo {year} {2018})}\BibitemShut {NoStop}%
\bibitem [{\citenamefont {Song}\ \emph {et~al.}(2019)\citenamefont {Song},
  \citenamefont {Yao},\ and\ \citenamefont {Wang}}]{song2019non}%
  \BibitemOpen
  \bibfield  {author} {\bibinfo {author} {\bibfnamefont {F.}~\bibnamefont
  {Song}}, \bibinfo {author} {\bibfnamefont {S.}~\bibnamefont {Yao}},\ and\
  \bibinfo {author} {\bibfnamefont {Z.}~\bibnamefont {Wang}},\ }\bibfield
  {title} {\bibinfo {title} {{Non-Hermitian skin effect and chiral damping in
  open quantum systems}},\ }\href
  {https://doi.org/10.1103/PhysRevLett.123.170401} {\bibfield  {journal}
  {\bibinfo  {journal} {Phys. Rev. Lett.}\ }\textbf {\bibinfo {volume} {123}},\
  \bibinfo {pages} {170401} (\bibinfo {year} {2019})}\BibitemShut {NoStop}%
\bibitem [{\citenamefont {Okuma}\ \emph {et~al.}(2020)\citenamefont {Okuma},
  \citenamefont {Kawabata}, \citenamefont {Shiozaki},\ and\ \citenamefont
  {Sato}}]{okuma2020topological}%
  \BibitemOpen
  \bibfield  {author} {\bibinfo {author} {\bibfnamefont {N.}~\bibnamefont
  {Okuma}}, \bibinfo {author} {\bibfnamefont {K.}~\bibnamefont {Kawabata}},
  \bibinfo {author} {\bibfnamefont {K.}~\bibnamefont {Shiozaki}},\ and\
  \bibinfo {author} {\bibfnamefont {M.}~\bibnamefont {Sato}},\ }\bibfield
  {title} {\bibinfo {title} {{Topological origin of non-Hermitian skin
  effects}},\ }\href {https://doi.org/10.1103/PhysRevLett.124.086801}
  {\bibfield  {journal} {\bibinfo  {journal} {Phys. Rev. Lett.}\ }\textbf
  {\bibinfo {volume} {124}},\ \bibinfo {pages} {086801} (\bibinfo {year}
  {2020})}\BibitemShut {NoStop}%
\bibitem [{\citenamefont {Kunst}\ \emph {et~al.}(2018)\citenamefont {Kunst},
  \citenamefont {Edvardsson}, \citenamefont {Budich},\ and\ \citenamefont
  {Bergholtz}}]{kunst2018biorthogonal}%
  \BibitemOpen
  \bibfield  {author} {\bibinfo {author} {\bibfnamefont {F.~K.}\ \bibnamefont
  {Kunst}}, \bibinfo {author} {\bibfnamefont {E.}~\bibnamefont {Edvardsson}},
  \bibinfo {author} {\bibfnamefont {J.~C.}\ \bibnamefont {Budich}},\ and\
  \bibinfo {author} {\bibfnamefont {E.~J.}\ \bibnamefont {Bergholtz}},\
  }\bibfield  {title} {\bibinfo {title} {Biorthogonal bulk-boundary
  correspondence in non-hermitian systems},\ }\href
  {https://doi.org/10.1103/PhysRevLett.121.026808} {\bibfield  {journal}
  {\bibinfo  {journal} {Phys. Rev. Lett.}\ }\textbf {\bibinfo {volume} {121}},\
  \bibinfo {pages} {026808} (\bibinfo {year} {2018})}\BibitemShut {NoStop}%
\bibitem [{\citenamefont {Berry}(2004)}]{berry2004physics}%
  \BibitemOpen
  \bibfield  {author} {\bibinfo {author} {\bibfnamefont {M.~V.}\ \bibnamefont
  {Berry}},\ }\bibfield  {title} {\bibinfo {title} {Physics of nonhermitian
  degeneracies},\ }\href {https://doi.org/10.1023/B:CJOP.0000044002.05657.04}
  {\bibfield  {journal} {\bibinfo  {journal} {Czechoslov. J. Phys.}\ }\textbf
  {\bibinfo {volume} {54}},\ \bibinfo {pages} {1039} (\bibinfo {year}
  {2004})}\BibitemShut {NoStop}%
\bibitem [{\citenamefont {Heiss}(2012)}]{heiss2012physics}%
  \BibitemOpen
  \bibfield  {author} {\bibinfo {author} {\bibfnamefont {W.}~\bibnamefont
  {Heiss}},\ }\bibfield  {title} {\bibinfo {title} {The physics of exceptional
  points},\ }\href {https://doi.org/10.1088/1751-8113/45/44/444016} {\bibfield
  {journal} {\bibinfo  {journal} {J. Phys. A: Math. Theor.}\ }\textbf {\bibinfo
  {volume} {45}},\ \bibinfo {pages} {444016} (\bibinfo {year}
  {2012})}\BibitemShut {NoStop}%
\bibitem [{\citenamefont {Kawabata}\ \emph
  {et~al.}(2019{\natexlab{a}})\citenamefont {Kawabata}, \citenamefont
  {Bessho},\ and\ \citenamefont {Sato}}]{kawabata2019classification}%
  \BibitemOpen
  \bibfield  {author} {\bibinfo {author} {\bibfnamefont {K.}~\bibnamefont
  {Kawabata}}, \bibinfo {author} {\bibfnamefont {T.}~\bibnamefont {Bessho}},\
  and\ \bibinfo {author} {\bibfnamefont {M.}~\bibnamefont {Sato}},\ }\bibfield
  {title} {\bibinfo {title} {{Classification of exceptional points and
  non-Hermitian topological semimetals}},\ }\href
  {https://doi.org/10.1103/PhysRevLett.123.066405} {\bibfield  {journal}
  {\bibinfo  {journal} {Phys. Rev. Lett.}\ }\textbf {\bibinfo {volume} {123}},\
  \bibinfo {pages} {066405} (\bibinfo {year} {2019}{\natexlab{a}})}\BibitemShut
  {NoStop}%
\bibitem [{\citenamefont {Gong}\ \emph {et~al.}(2018)\citenamefont {Gong},
  \citenamefont {Ashida}, \citenamefont {Kawabata}, \citenamefont {Takasan},
  \citenamefont {Higashikawa},\ and\ \citenamefont
  {Ueda}}]{gong2018topological}%
  \BibitemOpen
  \bibfield  {author} {\bibinfo {author} {\bibfnamefont {Z.}~\bibnamefont
  {Gong}}, \bibinfo {author} {\bibfnamefont {Y.}~\bibnamefont {Ashida}},
  \bibinfo {author} {\bibfnamefont {K.}~\bibnamefont {Kawabata}}, \bibinfo
  {author} {\bibfnamefont {K.}~\bibnamefont {Takasan}}, \bibinfo {author}
  {\bibfnamefont {S.}~\bibnamefont {Higashikawa}},\ and\ \bibinfo {author}
  {\bibfnamefont {M.}~\bibnamefont {Ueda}},\ }\bibfield  {title} {\bibinfo
  {title} {{Topological phases of non-Hermitian systems}},\ }\href
  {https://doi.org/10.1103/PhysRevX.8.031079} {\bibfield  {journal} {\bibinfo
  {journal} {Phys. Rev. X}\ }\textbf {\bibinfo {volume} {8}},\ \bibinfo {pages}
  {031079} (\bibinfo {year} {2018})}\BibitemShut {NoStop}%
\bibitem [{\citenamefont {Kawabata}\ \emph
  {et~al.}(2019{\natexlab{b}})\citenamefont {Kawabata}, \citenamefont
  {Shiozaki}, \citenamefont {Ueda},\ and\ \citenamefont
  {Sato}}]{kawabata2019symmetry}%
  \BibitemOpen
  \bibfield  {author} {\bibinfo {author} {\bibfnamefont {K.}~\bibnamefont
  {Kawabata}}, \bibinfo {author} {\bibfnamefont {K.}~\bibnamefont {Shiozaki}},
  \bibinfo {author} {\bibfnamefont {M.}~\bibnamefont {Ueda}},\ and\ \bibinfo
  {author} {\bibfnamefont {M.}~\bibnamefont {Sato}},\ }\bibfield  {title}
  {\bibinfo {title} {{Symmetry and topology in non-Hermitian physics}},\ }\href
  {https://doi.org/10.1103/PhysRevX.9.041015} {\bibfield  {journal} {\bibinfo
  {journal} {Phys. Rev. X}\ }\textbf {\bibinfo {volume} {9}},\ \bibinfo {pages}
  {041015} (\bibinfo {year} {2019}{\natexlab{b}})}\BibitemShut {NoStop}%
\bibitem [{\citenamefont {Jezequel}\ and\ \citenamefont
  {Delplace}(2023)}]{jezequel2023non}%
  \BibitemOpen
  \bibfield  {author} {\bibinfo {author} {\bibfnamefont {L.}~\bibnamefont
  {Jezequel}}\ and\ \bibinfo {author} {\bibfnamefont {P.}~\bibnamefont
  {Delplace}},\ }\bibfield  {title} {\bibinfo {title} {{Non-Hermitian spectral
  flows and Berry-Chern monopoles}},\ }\href
  {https://doi.org/10.1103/PhysRevLett.130.066601} {\bibfield  {journal}
  {\bibinfo  {journal} {Phys. Rev. Lett.}\ }\textbf {\bibinfo {volume} {130}},\
  \bibinfo {pages} {066601} (\bibinfo {year} {2023})}\BibitemShut {NoStop}%
\bibitem [{\citenamefont {Bender}\ and\ \citenamefont
  {Boettcher}(1998)}]{bender1998real}%
  \BibitemOpen
  \bibfield  {author} {\bibinfo {author} {\bibfnamefont {C.~M.}\ \bibnamefont
  {Bender}}\ and\ \bibinfo {author} {\bibfnamefont {S.}~\bibnamefont
  {Boettcher}},\ }\bibfield  {title} {\bibinfo {title} {{Real spectra in
  non-Hermitian Hamiltonians having P T symmetry}},\ }\href
  {https://doi.org/10.1103/PhysRevLett.80.5243} {\bibfield  {journal} {\bibinfo
   {journal} {Phys. Rev. Lett.}\ }\textbf {\bibinfo {volume} {80}},\ \bibinfo
  {pages} {5243} (\bibinfo {year} {1998})}\BibitemShut {NoStop}%
\bibitem [{\citenamefont {Bender}(2007)}]{bender2007making}%
  \BibitemOpen
  \bibfield  {author} {\bibinfo {author} {\bibfnamefont {C.~M.}\ \bibnamefont
  {Bender}},\ }\bibfield  {title} {\bibinfo {title} {{Making sense of
  non-Hermitian Hamiltonians}},\ }\href
  {https://doi.org/10.1088/0034-4885/70/6/R03} {\bibfield  {journal} {\bibinfo
  {journal} {Rep. Prog. Phys.}\ }\textbf {\bibinfo {volume} {70}},\ \bibinfo
  {pages} {947} (\bibinfo {year} {2007})}\BibitemShut {NoStop}%
\bibitem [{\citenamefont {El-Ganainy}\ \emph {et~al.}(2018)\citenamefont
  {El-Ganainy}, \citenamefont {Makris}, \citenamefont {Khajavikhan},
  \citenamefont {Musslimani}, \citenamefont {Rotter},\ and\ \citenamefont
  {Christodoulides}}]{el2018non}%
  \BibitemOpen
  \bibfield  {author} {\bibinfo {author} {\bibfnamefont {R.}~\bibnamefont
  {El-Ganainy}}, \bibinfo {author} {\bibfnamefont {K.~G.}\ \bibnamefont
  {Makris}}, \bibinfo {author} {\bibfnamefont {M.}~\bibnamefont {Khajavikhan}},
  \bibinfo {author} {\bibfnamefont {Z.~H.}\ \bibnamefont {Musslimani}},
  \bibinfo {author} {\bibfnamefont {S.}~\bibnamefont {Rotter}},\ and\ \bibinfo
  {author} {\bibfnamefont {D.~N.}\ \bibnamefont {Christodoulides}},\ }\bibfield
   {title} {\bibinfo {title} {{Non-Hermitian physics and PT symmetry}},\ }\href
  {https://doi.org/10.1038/nphys4323} {\bibfield  {journal} {\bibinfo
  {journal} {Nat. Phys.}\ }\textbf {\bibinfo {volume} {14}},\ \bibinfo {pages}
  {11} (\bibinfo {year} {2018})}\BibitemShut {NoStop}%
\bibitem [{\citenamefont {Zhang}\ \emph {et~al.}(2020)\citenamefont {Zhang},
  \citenamefont {Qin},\ and\ \citenamefont {Xiao}}]{Zhang2020}%
  \BibitemOpen
  \bibfield  {author} {\bibinfo {author} {\bibfnamefont {R.}~\bibnamefont
  {Zhang}}, \bibinfo {author} {\bibfnamefont {H.}~\bibnamefont {Qin}},\ and\
  \bibinfo {author} {\bibfnamefont {J.}~\bibnamefont {Xiao}},\ }\bibfield
  {title} {\bibinfo {title} {{{PT}-symmetry entails pseudo-Hermiticity
  regardless of diagonalizability}},\ }\href
  {https://doi.org/10.1063/1.5117211} {\bibfield  {journal} {\bibinfo
  {journal} {Journal of Mathematical Physics}\ }\textbf {\bibinfo {volume}
  {61}},\ \bibinfo {pages} {012101} (\bibinfo {year} {2020})}\BibitemShut
  {NoStop}%
\bibitem [{\citenamefont {Qin}\ \emph {et~al.}(2021)\citenamefont {Qin},
  \citenamefont {Fu}, \citenamefont {Glasser},\ and\ \citenamefont
  {Yahalom}}]{Qin2021}%
  \BibitemOpen
  \bibfield  {author} {\bibinfo {author} {\bibfnamefont {H.}~\bibnamefont
  {Qin}}, \bibinfo {author} {\bibfnamefont {Y.}~\bibnamefont {Fu}}, \bibinfo
  {author} {\bibfnamefont {A.~S.}\ \bibnamefont {Glasser}},\ and\ \bibinfo
  {author} {\bibfnamefont {A.}~\bibnamefont {Yahalom}},\ }\bibfield  {title}
  {\bibinfo {title} {Spontaneous and explicit parity-time-symmetry breaking in
  drift-wave instabilities},\ }\href
  {https://doi.org/10.1103/physreve.104.015215} {\bibfield  {journal} {\bibinfo
   {journal} {Physical Review E}\ }\textbf {\bibinfo {volume} {104}},\ \bibinfo
  {pages} {015215} (\bibinfo {year} {2021})}\BibitemShut {NoStop}%
\bibitem [{\citenamefont {Qin}\ \emph {et~al.}(2019)\citenamefont {Qin},
  \citenamefont {Zhang}, \citenamefont {Glasser},\ and\ \citenamefont
  {Xiao}}]{qin2019kelvin}%
  \BibitemOpen
  \bibfield  {author} {\bibinfo {author} {\bibfnamefont {H.}~\bibnamefont
  {Qin}}, \bibinfo {author} {\bibfnamefont {R.}~\bibnamefont {Zhang}}, \bibinfo
  {author} {\bibfnamefont {A.~S.}\ \bibnamefont {Glasser}},\ and\ \bibinfo
  {author} {\bibfnamefont {J.}~\bibnamefont {Xiao}},\ }\bibfield  {title}
  {\bibinfo {title} {{Kelvin-Helmholtz instability is the result of parity-time
  symmetry breaking}},\ }\href {https://doi.org/10.1063/1.5088498} {\bibfield
  {journal} {\bibinfo  {journal} {Phys. Plasmas}\ }\textbf {\bibinfo {volume}
  {26}},\ \bibinfo {pages} {032102} (\bibinfo {year} {2019})}\BibitemShut
  {NoStop}%
\bibitem [{\citenamefont {Fu}\ and\ \citenamefont {Qin}(2020)}]{fu2020physics}%
  \BibitemOpen
  \bibfield  {author} {\bibinfo {author} {\bibfnamefont {Y.}~\bibnamefont
  {Fu}}\ and\ \bibinfo {author} {\bibfnamefont {H.}~\bibnamefont {Qin}},\
  }\bibfield  {title} {\bibinfo {title} {{The physics of spontaneous
  parity-time symmetry breaking in the Kelvin--Helmholtz instability}},\ }\href
  {https://doi.org/10.1088/1367-2630/aba38f} {\bibfield  {journal} {\bibinfo
  {journal} {New J. Phys.}\ }\textbf {\bibinfo {volume} {22}},\ \bibinfo
  {pages} {083040} (\bibinfo {year} {2020})}\BibitemShut {NoStop}%
\bibitem [{\citenamefont {{\"O}zdemir}\ \emph {et~al.}(2019)\citenamefont
  {{\"O}zdemir}, \citenamefont {Rotter}, \citenamefont {Nori},\ and\
  \citenamefont {Yang}}]{ozdemir2019parity}%
  \BibitemOpen
  \bibfield  {author} {\bibinfo {author} {\bibfnamefont {{\c{S}}.~K.}\
  \bibnamefont {{\"O}zdemir}}, \bibinfo {author} {\bibfnamefont
  {S.}~\bibnamefont {Rotter}}, \bibinfo {author} {\bibfnamefont
  {F.}~\bibnamefont {Nori}},\ and\ \bibinfo {author} {\bibfnamefont
  {L.}~\bibnamefont {Yang}},\ }\bibfield  {title} {\bibinfo {title}
  {Parity--time symmetry and exceptional points in photonics},\ }\href
  {https://doi.org/10.1038/nphys4323} {\bibfield  {journal} {\bibinfo
  {journal} {Nat. materials}\ }\textbf {\bibinfo {volume} {18}},\ \bibinfo
  {pages} {783} (\bibinfo {year} {2019})}\BibitemShut {NoStop}%
\bibitem [{\citenamefont {Miri}\ and\ \citenamefont
  {Alu}(2019)}]{miri2019exceptional}%
  \BibitemOpen
  \bibfield  {author} {\bibinfo {author} {\bibfnamefont {M.-A.}\ \bibnamefont
  {Miri}}\ and\ \bibinfo {author} {\bibfnamefont {A.}~\bibnamefont {Alu}},\
  }\bibfield  {title} {\bibinfo {title} {Exceptional points in optics and
  photonics},\ }\href {https://doi.org/10.1126/science.aar7709} {\bibfield
  {journal} {\bibinfo  {journal} {Science}\ }\textbf {\bibinfo {volume}
  {363}},\ \bibinfo {pages} {eaar7709} (\bibinfo {year} {2019})}\BibitemShut
  {NoStop}%
\bibitem [{\citenamefont {Moyal}(1949)}]{moyal1949quantum}%
  \BibitemOpen
  \bibfield  {author} {\bibinfo {author} {\bibfnamefont {J.~E.}\ \bibnamefont
  {Moyal}},\ }\bibfield  {title} {\bibinfo {title} {Quantum mechanics as a
  statistical theory},\ }\href {https://doi.org/10.1017/S0305004100000487}
  {\bibfield  {journal} {\bibinfo  {journal} {Math. Proc. Camb.}\ }\textbf
  {\bibinfo {volume} {45}},\ \bibinfo {pages} {99} (\bibinfo {year}
  {1949})}\BibitemShut {NoStop}%
\bibitem [{\citenamefont {Hillery}\ \emph {et~al.}(1984)\citenamefont
  {Hillery}, \citenamefont {O'Connell}, \citenamefont {Scully},\ and\
  \citenamefont {Wigner}}]{hillery1984distribution}%
  \BibitemOpen
  \bibfield  {author} {\bibinfo {author} {\bibfnamefont {M.}~\bibnamefont
  {Hillery}}, \bibinfo {author} {\bibfnamefont {R.}~\bibnamefont {O'Connell}},
  \bibinfo {author} {\bibfnamefont {M.}~\bibnamefont {Scully}},\ and\ \bibinfo
  {author} {\bibfnamefont {E.}~\bibnamefont {Wigner}},\ }\bibfield  {title}
  {\bibinfo {title} {Distribution functions in physics: Fundamentals},\ }\href
  {https://doi.org/https://doi.org/10.1016/0370-1573(84)90160-1} {\bibfield
  {journal} {\bibinfo  {journal} {Phys. Rep.}\ }\textbf {\bibinfo {volume}
  {106}},\ \bibinfo {pages} {121} (\bibinfo {year} {1984})}\BibitemShut
  {NoStop}%
\bibitem [{\citenamefont {Xu}\ \emph {et~al.}(2017)\citenamefont {Xu},
  \citenamefont {Wang},\ and\ \citenamefont {Duan}}]{xu2017weyl}%
  \BibitemOpen
  \bibfield  {author} {\bibinfo {author} {\bibfnamefont {Y.}~\bibnamefont
  {Xu}}, \bibinfo {author} {\bibfnamefont {S.-T.}\ \bibnamefont {Wang}},\ and\
  \bibinfo {author} {\bibfnamefont {L.-M.}\ \bibnamefont {Duan}},\ }\bibfield
  {title} {\bibinfo {title} {Weyl exceptional rings in a three-dimensional
  dissipative cold atomic gas},\ }\href
  {https://doi.org/10.1103/PhysRevLett.118.045701} {\bibfield  {journal}
  {\bibinfo  {journal} {Phys. Rev. Lett.}\ }\textbf {\bibinfo {volume} {118}},\
  \bibinfo {pages} {045701} (\bibinfo {year} {2017})}\BibitemShut {NoStop}%
\bibitem [{\citenamefont {Cerjan}\ \emph {et~al.}(2018)\citenamefont {Cerjan},
  \citenamefont {Xiao}, \citenamefont {Yuan},\ and\ \citenamefont
  {Fan}}]{cerjan2018effects}%
  \BibitemOpen
  \bibfield  {author} {\bibinfo {author} {\bibfnamefont {A.}~\bibnamefont
  {Cerjan}}, \bibinfo {author} {\bibfnamefont {M.}~\bibnamefont {Xiao}},
  \bibinfo {author} {\bibfnamefont {L.}~\bibnamefont {Yuan}},\ and\ \bibinfo
  {author} {\bibfnamefont {S.}~\bibnamefont {Fan}},\ }\bibfield  {title}
  {\bibinfo {title} {{Effects of non-Hermitian perturbations on Weyl
  Hamiltonians with arbitrary topological charges}},\ }\href
  {https://doi.org/10.1103/PhysRevB.97.075128} {\bibfield  {journal} {\bibinfo
  {journal} {Phys. Rev. B}\ }\textbf {\bibinfo {volume} {97}},\ \bibinfo
  {pages} {075128} (\bibinfo {year} {2018})}\BibitemShut {NoStop}%
\bibitem [{\citenamefont {Cerjan}\ \emph {et~al.}(2019)\citenamefont {Cerjan},
  \citenamefont {Huang}, \citenamefont {Wang}, \citenamefont {Chen},
  \citenamefont {Chong},\ and\ \citenamefont
  {Rechtsman}}]{cerjan2019experimental}%
  \BibitemOpen
  \bibfield  {author} {\bibinfo {author} {\bibfnamefont {A.}~\bibnamefont
  {Cerjan}}, \bibinfo {author} {\bibfnamefont {S.}~\bibnamefont {Huang}},
  \bibinfo {author} {\bibfnamefont {M.}~\bibnamefont {Wang}}, \bibinfo {author}
  {\bibfnamefont {K.~P.}\ \bibnamefont {Chen}}, \bibinfo {author}
  {\bibfnamefont {Y.}~\bibnamefont {Chong}},\ and\ \bibinfo {author}
  {\bibfnamefont {M.~C.}\ \bibnamefont {Rechtsman}},\ }\bibfield  {title}
  {\bibinfo {title} {{Experimental realization of a Weyl exceptional ring}},\
  }\href {https://doi.org/10.1038/s41566-019-0453-z} {\bibfield  {journal}
  {\bibinfo  {journal} {Nat. Photonics}\ }\textbf {\bibinfo {volume} {13}},\
  \bibinfo {pages} {623} (\bibinfo {year} {2019})}\BibitemShut {NoStop}%
\bibitem [{\citenamefont {Beekman}\ \emph {et~al.}(2019)\citenamefont
  {Beekman}, \citenamefont {Rademaker},\ and\ \citenamefont {van
  Wezel}}]{beekman2019introduction}%
  \BibitemOpen
  \bibfield  {author} {\bibinfo {author} {\bibfnamefont {A.~J.}\ \bibnamefont
  {Beekman}}, \bibinfo {author} {\bibfnamefont {L.}~\bibnamefont {Rademaker}},\
  and\ \bibinfo {author} {\bibfnamefont {J.}~\bibnamefont {van Wezel}},\
  }\bibfield  {title} {\bibinfo {title} {{An introduction to spontaneous
  symmetry breaking}},\ }\href
  {https://doi.org/10.21468/SciPostPhysLectNotes.11} {\bibfield  {journal}
  {\bibinfo  {journal} {SciPost Phys. Lect. Notes}\ ,\ \bibinfo {pages} {11}}
  (\bibinfo {year} {2019})}\BibitemShut {NoStop}%
\bibitem [{Note1()}]{Note1}%
  \BibitemOpen
  \bibinfo {note} {Some antilinear operators do not have eigenvectors. For
  instance, with $\protect \mathcal {P}=\protect \mathrm {i}\sigma ^{y}$,
  $\protect \mathcal {PT}$ has no eigenvector because $(\protect \mathcal
  {PT})^{2}=-1$. However, since classical systems have $(\protect \mathcal
  {PT})^{2}=1$, this issue is avoided. See Ref.~\cite {uhlmann2016anti} for
  example.}\BibitemShut {Stop}%
\bibitem [{Note2()}]{Note2}%
  \BibitemOpen
  \bibinfo {note} {It is equivalent to restricting our discussion on type-I
  Weyl points defined in Ref.\cite {soluyanov2015type}.}\BibitemShut {Stop}%
\bibitem [{\citenamefont {Goedbloed}\ and\ \citenamefont
  {Poedts}(2004)}]{goedbloed_poedts_2004}%
  \BibitemOpen
  \bibfield  {author} {\bibinfo {author} {\bibfnamefont {J.~P.~H.}\
  \bibnamefont {Goedbloed}}\ and\ \bibinfo {author} {\bibfnamefont
  {S.}~\bibnamefont {Poedts}},\ }\href
  {https://doi.org/10.1017/CBO9780511616945} {\emph {\bibinfo {title}
  {{Principles of Magnetohydrodynamics}}}}\ (\bibinfo  {publisher} {Cambridge
  University Press},\ \bibinfo {year} {2004})\BibitemShut {NoStop}%
\bibitem [{\citenamefont {Freidberg}(2014)}]{freidberg_2014}%
  \BibitemOpen
  \bibfield  {author} {\bibinfo {author} {\bibfnamefont {J.~P.}\ \bibnamefont
  {Freidberg}},\ }\href {https://doi.org/10.1017/CBO9780511795046} {\emph
  {\bibinfo {title} {Ideal MHD}}}\ (\bibinfo  {publisher} {Cambridge University
  Press},\ \bibinfo {year} {2014})\BibitemShut {NoStop}%
\bibitem [{\citenamefont {Hameiri}\ and\ \citenamefont
  {Torasso}(2004)}]{hameiri2004linear}%
  \BibitemOpen
  \bibfield  {author} {\bibinfo {author} {\bibfnamefont {E.}~\bibnamefont
  {Hameiri}}\ and\ \bibinfo {author} {\bibfnamefont {R.}~\bibnamefont
  {Torasso}},\ }\bibfield  {title} {\bibinfo {title} {{Linear stability of
  static equilibrium states in the Hall-magnetohydrodynamics model}},\ }\href
  {https://doi.org/10.1063/1.1784453} {\bibfield  {journal} {\bibinfo
  {journal} {Phys. Plasmas}\ }\textbf {\bibinfo {volume} {11}},\ \bibinfo
  {pages} {4934} (\bibinfo {year} {2004})}\BibitemShut {NoStop}%
\bibitem [{\citenamefont {Hameiri}\ \emph {et~al.}(2005)\citenamefont
  {Hameiri}, \citenamefont {Ishizawa},\ and\ \citenamefont
  {Ishida}}]{hameiri2005waves}%
  \BibitemOpen
  \bibfield  {author} {\bibinfo {author} {\bibfnamefont {E.}~\bibnamefont
  {Hameiri}}, \bibinfo {author} {\bibfnamefont {A.}~\bibnamefont {Ishizawa}},\
  and\ \bibinfo {author} {\bibfnamefont {A.}~\bibnamefont {Ishida}},\
  }\bibfield  {title} {\bibinfo {title} {{Waves in the
  Hall-magnetohydrodynamics model}},\ }\href
  {https://doi.org/10.1063/1.1952887} {\bibfield  {journal} {\bibinfo
  {journal} {Phys. Plasmas}\ }\textbf {\bibinfo {volume} {12}},\ \bibinfo
  {pages} {072109} (\bibinfo {year} {2005})}\BibitemShut {NoStop}%
\bibitem [{\citenamefont {Uhlmann}(2016)}]{uhlmann2016anti}%
  \BibitemOpen
  \bibfield  {author} {\bibinfo {author} {\bibfnamefont {A.}~\bibnamefont
  {Uhlmann}},\ }\bibfield  {title} {\bibinfo {title} {Anti-(conjugate)
  linearity},\ }\href {https://doi.org/10.1007/s11433-015-5777-1} {\bibfield
  {journal} {\bibinfo  {journal} {Sci. China Phys. Mech. Astron.}\ }\textbf
  {\bibinfo {volume} {59}},\ \bibinfo {pages} {1} (\bibinfo {year}
  {2016})}\BibitemShut {NoStop}%
\bibitem [{\citenamefont {Soluyanov}\ \emph {et~al.}(2015)\citenamefont
  {Soluyanov}, \citenamefont {Gresch}, \citenamefont {Wang}, \citenamefont
  {Wu}, \citenamefont {Troyer}, \citenamefont {Dai},\ and\ \citenamefont
  {Bernevig}}]{soluyanov2015type}%
  \BibitemOpen
  \bibfield  {author} {\bibinfo {author} {\bibfnamefont {A.~A.}\ \bibnamefont
  {Soluyanov}}, \bibinfo {author} {\bibfnamefont {D.}~\bibnamefont {Gresch}},
  \bibinfo {author} {\bibfnamefont {Z.}~\bibnamefont {Wang}}, \bibinfo {author}
  {\bibfnamefont {Q.}~\bibnamefont {Wu}}, \bibinfo {author} {\bibfnamefont
  {M.}~\bibnamefont {Troyer}}, \bibinfo {author} {\bibfnamefont
  {X.}~\bibnamefont {Dai}},\ and\ \bibinfo {author} {\bibfnamefont {B.~A.}\
  \bibnamefont {Bernevig}},\ }\bibfield  {title} {\bibinfo {title} {{Type-II
  Weyl semimetals}},\ }\href {https://doi.org/10.1038/nature15768} {\bibfield
  {journal} {\bibinfo  {journal} {Nature}\ }\textbf {\bibinfo {volume} {527}},\
  \bibinfo {pages} {495} (\bibinfo {year} {2015})}\BibitemShut {NoStop}%
\end{thebibliography}%

\end{document}